\documentclass[aps,prl,twocolumn,showpacs,longbibliography]{revtex4-1}

\usepackage{amssymb,amsmath,multirow,hyperref,txfonts,amsfonts,array,color}
\usepackage[pdftex]{graphicx}
\usepackage[margin=1.0in]{geometry}
\usepackage[locale=US,output-product=\cdot,]{siunitx}

\newcommand{\cb}[1]{\left\{#1\right\}}

\newcommand{\fig}{Fig.}
 
\newcommand{\Sexp}{S_{\rm exp}}
\newcommand{\sepr}{\vec s}
\newcommand{\qaa}{{\hat{n}_{S_{A}}}}
\newcommand{\qbb}{{\hat{n}_{S_{B}}}}
\newcommand{\raa}{\vec r_A}
\newcommand{\rbb}{\vec r_B}
\newcommand{\maa}{\vec m_A}
\newcommand{\mbb}{\vec m_B}

\newcommand{\na}{\eta_A} 
\newcommand{\nb}{\eta_B}
\newcommand{\nair}{n}

\newcommand{\skips}{-0.4cm}
\newcommand{\vskips}{-0.5cm}

\newcommand{\pasp}{Pub. Astr. Soc. Pacific}
\newcommand{\aap}{  Astron. \& Astrophys. }
\begin{document}
%------------------------------------------------------------------------------------------------------------------------------------------------

\title{Cosmic Bell Test: Measurement Settings from Milky Way Stars}

\author{Johannes Handsteiner$^1$}
\email{johannes.handsteiner@univie.ac.at}
\author{Andrew S. Friedman$^2$}
\email{asf@mit.edu}
\author{Dominik Rauch$^1$,
Jason Gallicchio$^3$,
Bo Liu$^{1,4}$,
Hannes Hosp$^1$,
Johannes Kofler$^5$,
David Bricher$^1$,
Matthias Fink$^1$,
Calvin Leung$^3$,
Anthony Mark$^2$,
Hien T. Nguyen$^6$,
Isabella Sanders$^2$,
Fabian Steinlechner$^1$,
Rupert Ursin$^{1,7}$,
S{\"o}ren Wengerowsky$^1$,
Alan H. Guth$^2$,
David I. Kaiser$^2$,
Thomas Scheidl$^1$}
\author{Anton Zeilinger$^{1,7}$}\email{anton.zeilinger@univie.ac.at}

\affiliation{
$^1$Institute for Quantum Optics and Quantum Information (IQOQI), Austrian Academy of Sciences, Boltzmanngasse 3, 1090 Vienna, Austria \\
$^2$Department of Physics, Massachusetts Institute of Technology, Cambridge, MA 02139 USA \\
$^3$Department of Physics, Harvey Mudd College, Claremont, CA 91711 USA \\
$^4$School of Computer, NUDT, 410073 Changsha, China \\
$^5$Max Planck Institute of Quantum Optics, Hans-Kopfermann-Stra{\ss}e 1, 85748 Garching, Germany \\
$^6$NASA Jet Propulsion Laboratory, Pasadena, CA 91109 USA\\
$^7$Vienna Center for Quantum Science \& Technology (VCQ), Faculty of Physics, University of Vienna, Boltzmanngasse 5, 1090 Vienna, Austria
}

\date{\today}

%------------------------------------------------------------------------------------------------------------------------------------------------

\begin{abstract} 

Bell's theorem states that some predictions of quantum mechanics cannot be reproduced by a local-realist theory. That conflict is expressed by Bell's inequality, which is usually derived under the assumption that there are no statistical correlations between the choices of measurement settings and anything else that can causally affect the measurement outcomes. In previous experiments, this ``freedom of choice" was addressed by ensuring that selection of measurement settings via conventional ``quantum random number generators" was space-like separated from the entangled particle creation. This, however, left open the possibility that an unknown cause affected both the setting choices and measurement outcomes as recently as mere microseconds before each experimental trial. Here we report on a new experimental test of Bell's inequality that, for the first time, uses distant astronomical sources as ``cosmic setting generators." In our tests with polarization-entangled photons, measurement settings were chosen using real-time observations of Milky Way stars while simultaneously ensuring locality. 
Assuming fair sampling for all detected photons, and that each stellar photon's color was set at emission, we
observe statistically significant $\gtrsim 7.31 \sigma$ and $\gtrsim 11.93 \sigma$ violations of Bell's inequality with estimated $p$-values of $ \lesssim 1.8 \times 10^{-13}$ and $\lesssim 4.0 \times 10^{-33}$, respectively, thereby pushing back by $\sim$600 years the most recent time by which any local-realist influences could have engineered the observed Bell violation.
\end{abstract}

%\pacs{03.65.Ud, 42.50.Xa, 97.20.-w, 95.55.Qf, 95.75.De;  Preprint MIT-CTP 4854} 

\maketitle

%-----------------------------------------------------------------------------------------------------------------------------------------------

In this Letter, we report on a new experimental test of Bell's inequality that, for the first time, uses distant astronomical sources to choose measurement settings. 
This is the first in a series of ``cosmic Bell tests" that will use progressively more distant sources, ultimately pushing the measurement settings' origin to greater and greater cosmological distances \cite{gallicchio14}.

%%%%%%%%%%%%%%%%%%%%%%%%%%%%%%%%%%%%%%%%
\begin{figure*}[!htp]
\scriptsize
\includegraphics[width=5.9in]{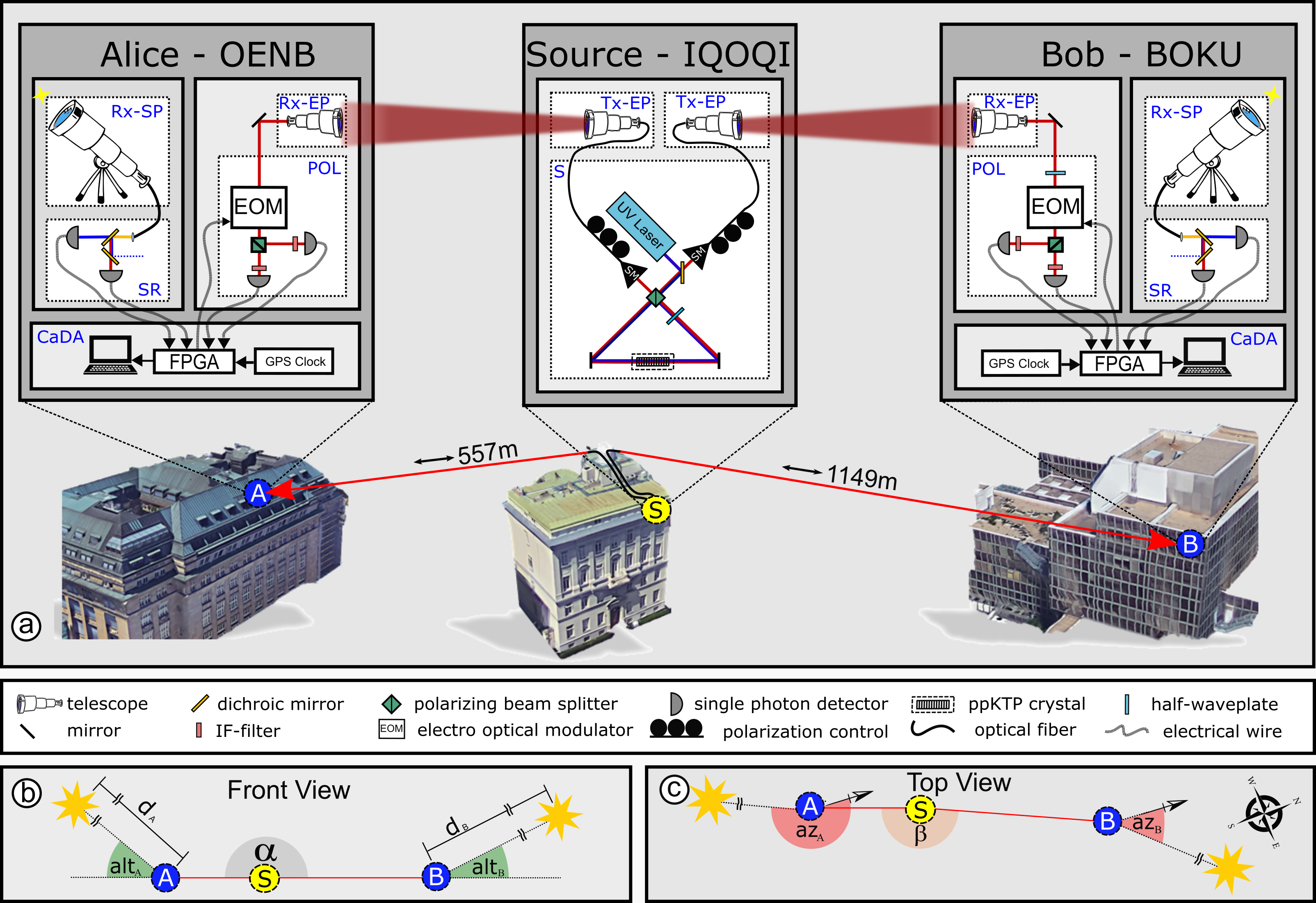}
\vspace{-0.4cm}
\caption{
\small
(a) The three experimental stations and related acronyms are described in the main text. Stellar photon receiving telescopes (Rx-SP) with primary mirror diameters of 0.2032\,m (Meade 8-inch LX200 ACF, $f$=2\,m) and 0.254\,m (Meade 10-inch LX600 ACF, $f$=2.032\,m) were used by Alice and Bob, respectively, although the telescope apertures were each partially covered to limit sky noise. 
Diameters and focal lengths of the quantum channel telescopes are: Alice (Tx-EP: $d$=50.8\,mm, $f$=100\,mm; Rx-EP: $d$=80\,mm, $f$=400\,mm) and Bob (Tx-EP: $d$=70\,mm, $f$=280\,mm; Rx-EP: $d$=140\,mm, $f$=420\,mm).
Latitude, longitude, and elevation for the three experimental sites are: Alice ({\it A}: 48.21645$^{\circ}$, 16.354311$^{\circ}$, 215.0\, m), Bob ({\it B}: 48.23160$^{\circ}$, 16.3579553$^{\circ}$, 200.0\,m), and the Source ({\it S}: 48.221311$^{\circ}$, 16.356439$^{\circ}$, 205.0\,m). (b) and (c) For experimental run 1, the setup deviated from the ideal, co-linear 1D case by the angles displayed. Site coordinates yield $\alpha \approx 180^{\circ}$ and $\beta \approx 169^{\circ}$ for both runs. See Table~\ref{tab:exp} for the azimuth (az) and altitude (alt) of each star observed during each experimental run. (3D graphics taken from Google Earth, 2016.) 
\vspace{-0.4cm}
}
\label{fig:sitemap}
\end{figure*}
%%%%%%%%%%%%%%%%%%%%%%%%%%%%%%%%%%%%%%%%

{\it Background.---} 
Scientists have struggled with the alleged incompatibility of quantum entanglement and our everyday intuitions about the physical world since the seminal paper by Einstein, Podolsky and Rosen (EPR) in 1935 \cite{einstein35}. EPR concluded that the description of reality given by the quantum-mechanical wave-function is incomplete because it is incompatible with the concepts of {\it locality} (no physical influences can travel faster than the speed of light in vacuum) and {\it realism}  
(objects possess complete sets of properties on their own, prior to measurement).
A well-known ``tool" to experimentally distinguish between the quantum predictions and local-realist alternatives of the sort envisaged by EPR is provided by the famous inequality derived by John Bell in 1964 \cite{bell64}. Assuming locality and realism, Bell's inequality limits the degree to which measurement outcomes on pairs of distant systems may be correlated, if the measurements of one system are carried out with only limited information about the other. By contrast, measurements on entangled particle pairs in the quantum singlet state, for example, are predicted to violate Bell's inequality. Beginning with \cite{freedman72}, essentially all significant experimental Bell tests to date have supported the quantum-mechanical predictions.

However, the conclusions of any experiment are valid only given certain assumptions, the violation of which leaves open ``loopholes," whereby a local-realist description of nature could still be compatible with the experimental results. (For extensive reviews of Bell-test loopholes, see \cite{brunner14,larsson14b,kofler16}.) For example, the locality loophole concerns whether any information about one side's measurement setting or measurement outcome could have been communicated (at or below the speed of light) to the other side prior to its measurement. This loophole has been closed by space-like separating measurement-setting choices on each side from the other side's measurement outcomes \cite{aspect82,weihs98}. The fair-sampling loophole \cite{pearle70} concerns whether the set of entangled particles detected on both sides was representative of all emitted pairs rather than a biased sub-ensemble, and has recently been closed by ensuring a sufficient total fraction of detected pairs  \cite{rowe01,matsukevich08,ansmann09,hofmann12,giustina13,christensen13}. Even more recently, several cutting-edge experiments have demonstrated violations of Bell's inequality while closing both the locality and fair-sampling loopholes simultaneously \cite{hensen15,giustina15,shalm15,hensen16,rosenfeld16}.

A third major loophole, known variously as the freedom-of-choice, measurement-independence, or setting-independence loophole \cite{bell76,shimony76,bell77,brans88}, concerns the choice of measurement settings. In particular, the derivation of Bell's inequality explicitly assumes that there is no statistical correlation between the choices of measurement settings and anything else that causally affects both measurement outcomes. Bell himself observed forty years ago that, ``It has been assumed that the settings of instruments are in some sense free variables---say at the whim of experimenters---or in any case not determined in the overlap of the backward light cones" \cite{bell76}. Recent theoretical work has demonstrated that models that relax this assumption, allowing for a modest correlation between the joint measurement settings and any causal
influence on the measurement outcomes, can reproduce the quantum correlations \cite{kofler06,hall10,hall11,barrett11,koh12,banik12,colbeck12,thinh13,putz14,putz16,hall16}. 
(See also \cite{pironio15} on some subtleties of addressing the freedom-of-choice loophole.)

Even if nature does not exploit this loophole, testing it experimentally (e.g. \cite{scheidl10,aktas15}) has significant practical relevance for device-independent quantum key distribution \cite{barrett05,pironio09,vazirani14} as well as random-number generation and randomness expansion \cite{pironio10,colbeck12,gallego13}.
In particular, a sophisticated adversary could undermine a variety of
quantum information schemes by utilizing the freedom-of-choice loophole \cite{kofler06,koh12,hall16}.

To the extent that recent experiments have addressed freedom of choice, they have adopted the additional, strong assumption that the relevant causal influences (or ``hidden variables") originate together with the entangled particles and hence cannot influence setting choices in space-like separated regions \cite{scheidl10,giustina15,shalm15}, or assumed that the setting-generation process is completely independent of its past \cite{hensen15,hensen16,rosenfeld16}. Yet nowhere in the derivation of Bell's inequality does the formalism make any stipulation about where or when such hidden variables could be created or become relevant. In fact, as Bell himself emphasized \cite{bell76,bell77} (see also \cite{colbeck12,gallicchio14}), they could be associated with any events within the experiment's past light-cone \cite{Retrofn}. Thus, in principle, the possibility exists that the purportedly random setting choices in previous experiments could have been influenced by some unknown cause in their past, exploiting the freedom-of-choice loophole through events as recent as a few microseconds before the measurements \cite{Superdeterminism}.

Here we report on an experimental Bell test with polarization entangled photons that, assuming fair sampling, significantly constrains the space-time region from which any such unknown, causal influences could have affected both the measurement settings and outcomes. While simultaneously closing the locality loophole, measurement settings are determined by ``cosmic setting generators," 
using the color of photons detected during real-time astronomical observations of distant stars. 
We observed Bell violations with high statistical significance and thus conclude that any hidden causal influences that could have exploited the freedom-of-choice loophole would have to have originated from remote space-time events at least several hundred years ago, at locations seemingly unrelated to the entangled-pair creation. Compared to previous experiments, this pushes back by $\sim$16 orders of magnitude the most recent time by which any local-realist influences could have engineered the observed correlations.

The idea to address freedom-of-choice by using distant astronomical sources to choose Bell-test measurement settings was already discussed as far back as the 1976 Erice meeting organized by John Bell and Bernard d'Espagnat \cite{erice76}. While others have also briefly noted this basic premise 
\cite{maudlin94,vaidman01,scheidl10}, this work is the first to implement it experimentally, building on a detailed feasibility study \cite{gallicchio14}.

%-----------------------------------------------------------------------------------------------------------------------------------------------

{\it Experimental Implementation.---}  Fig.~\ref{fig:sitemap} shows our three experimental sites across Vienna. A central entangled photon source $S$ was located in a laboratory on the 4th floor of the Institute for Quantum Optics and Quantum Information (IQOQI) and the two observers, Alice (A) and Bob (B), were situated on the 9th floor of the Austrian National Bank (OENB) and on the 5th floor of the University of Natural Resources and Life Sciences (BOKU), respectively. 

The entangled photon source is based on a Sagnac interferometer \cite{kim06, fedrizzi07} generating polarization-entangled photon pairs in the maximally entangled singlet state $\left| \Psi^{-} \right> \!=\! \frac{1}{\sqrt{2}}( \left| H_A V_B \right> - \left| V_A H_B \right>)$. Using single-mode fibers, each entangled photon was guided to an entangled photon transmitting telescope (Tx-EP) located at the rooftop of IQOQI, which sent the photons via free-space quantum channels to Alice and Bob, respectively. Measurement stations for Alice and Bob each featured an entangled photon receiving telescope (Rx-EP), a polarization analyzer (POL), stellar 
photon receiving telescope (Rx-SP), a setting reader (SR) and a control and data acquisition unit (CaDA). The entangled photons were collected with the Rx-EP and guided to the polarization analyzer where an electro-optical modulator (EOM) allowed for fast switching between complementary measurement bases. This was followed by a polarizing beam splitter with a single-photon avalanche diode (SPAD) detector in each output port.

The Rx-SPs collected stellar photons, which were guided by multimode fibers to setting readers, where dichroic mirrors with $\sim$700\,nm cutoffs split them into ``blue" and ``red" arms, each fed to a SPAD. An FPGA board processed the SPAD signals to electronically implement the corresponding EOM measurement setting. Every detector click in a red/blue arm induced a measurement in the following linear polarization bases for Alice: 45$^\circ$/135$^\circ$ (blue) and 0$^\circ$/90$^\circ$ (red), and Bob: 22.5$^\circ$/112.5$^\circ$ (blue) and -22.5$^\circ$/67.5$^\circ$ (red), respectively. Finally, using a GPS-disciplined clock, all SPAD detections from the polarization analyzer and the setting reader were time-stamped by the FPGA board and recorded by a computer.

We specifically use photon color to implement measurement settings under the assumption that the wavelength of the photon emitted from the star was determined at the time of emission and unaltered since. Astrophysical motivations for this assumption include the absence of any known mechanism that preferentially re-radiates photons at a different wavelength along our line of sight;
any such process would violate the conservation of energy and momentum. 
In addition, the effects of wavelength-dependent attenuation by the interstellar medium are negligible for Milky Way stars within a few thousand light years (ly) \cite{cardelli89,fitzpatrick99,schlafly11}. By contrast, significant attenuation  from the Earth's atmosphere (38-45\% loss) as well as from the experimental setup (59-61\% loss) is unavoidable (see the Supplemental Material \cite{SM-ref}). Thus, our approach requires the assumption that this represents a fair sample of the celestial emissions.
 
Two other major effects must be considered to account for SPAD detections which do not represent stellar photons with correctly identified colors. First, local sources of noise, including sky glow, light pollution, and dark counts determine between 1\% and 5\% of the settings, as measured by looking at a dark patch of sky next to each star. Only a tenth of these are detector dark counts. In addition, due to imperfect dichroic mirrors, a certain fraction of detected stellar photons will be assigned the wrong setting. Hence, to precisely quantify the full optical path of our settings implementation method, we measured transmission and reflection spectra for the dichroic mirrors, and multiplied this by the spectra of the stars, atmospheric transmission \cite{berk1987modtran}, reflection and transmission of optical elements, and the SPAD detector response. Conservative estimates of the fraction of incorrectly determined measurement settings range from $f_w \approx$$1.4$-$2.0$\%, depending on the stellar spectrum and red/blue output port (see the Supplemental Material \cite{SM-ref}).

%%%%%%%%%%%%%%%%%%%%%%%%%%%%%%%%%%%%%%%%
\begin{figure}
\centering
\begin{tabular}{@{}c@{}}
\includegraphics[width=2.9in]{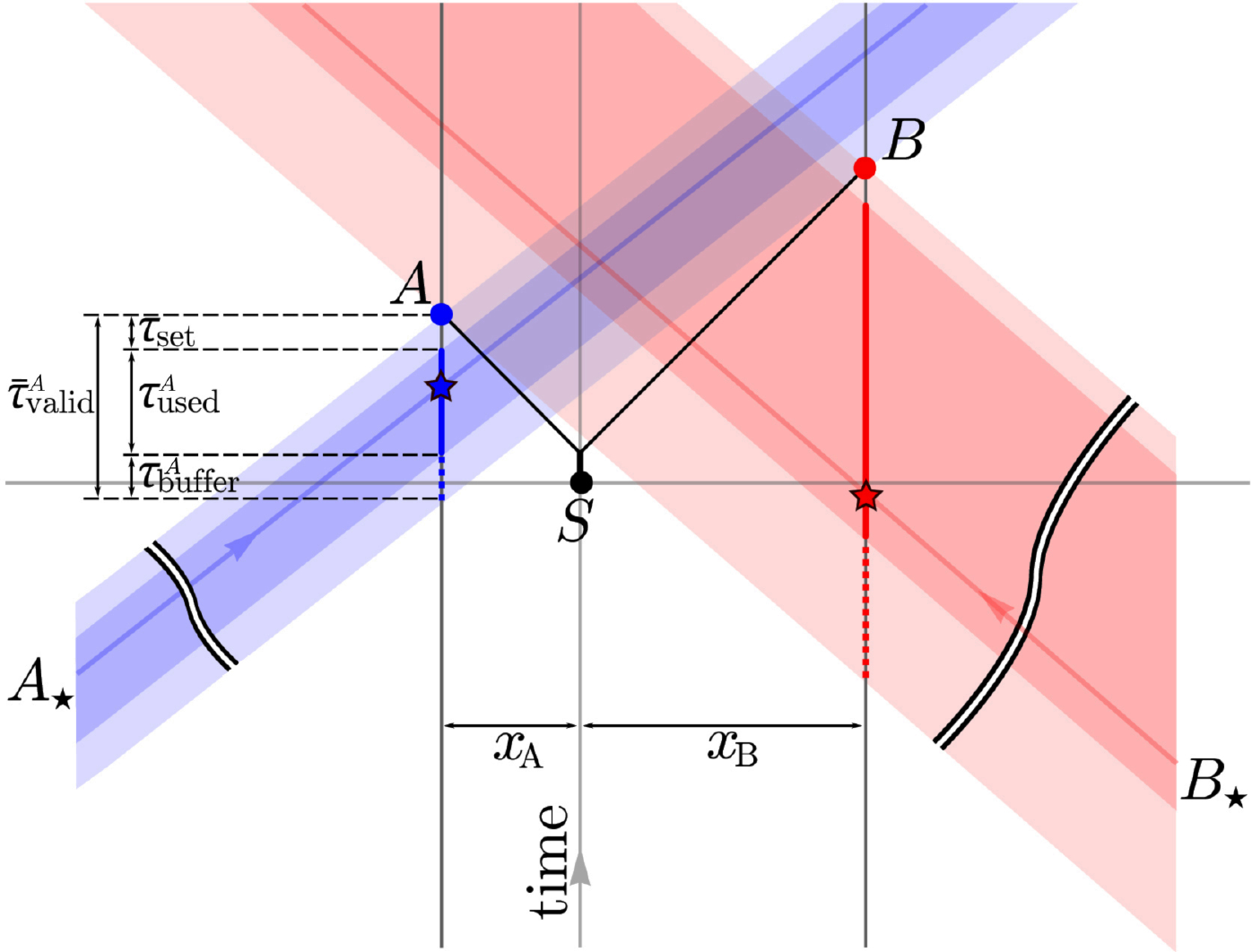} 
\end{tabular}
\vspace{-0.4cm}
\caption{
\small
1+1\,D space-time diagram for run 1, with the origin at the entangled pair creation (black dot) and a spatial projection axis chosen to minimize its distance to Alice and Bob. After a fiber delay (thick black line), %EPR 
entangled photons are sent via free-space channels (thin black lines) to be measured by Alice and Bob at events $A$ and $B$. Blue and red stars indicate example valid settings from measuring stellar photons emitted far away at space-time events $A_{\star}$ and $B_{\star}$ (see Fig.~\ref{fig:space}). Ensuring locality limits valid settings to the shaded regions. Delays to implement each setting and an added safety buffer shorten the validity time windows actually used to the darker shaded regions.
\vspace{-0.4cm}
}
\label{fig:space2}
\end{figure}
%%%%%%%%%%%%%%%%%%%%%%%%%%%%%%%%%%%%%%%%

%%%%%%%%%%%%%%%%%%%%%%%%%%%%%%%%%%%%%%%%
\begin{figure}
\centering
\begin{tabular}{@{}c@{}}
\includegraphics[width=2.8in]{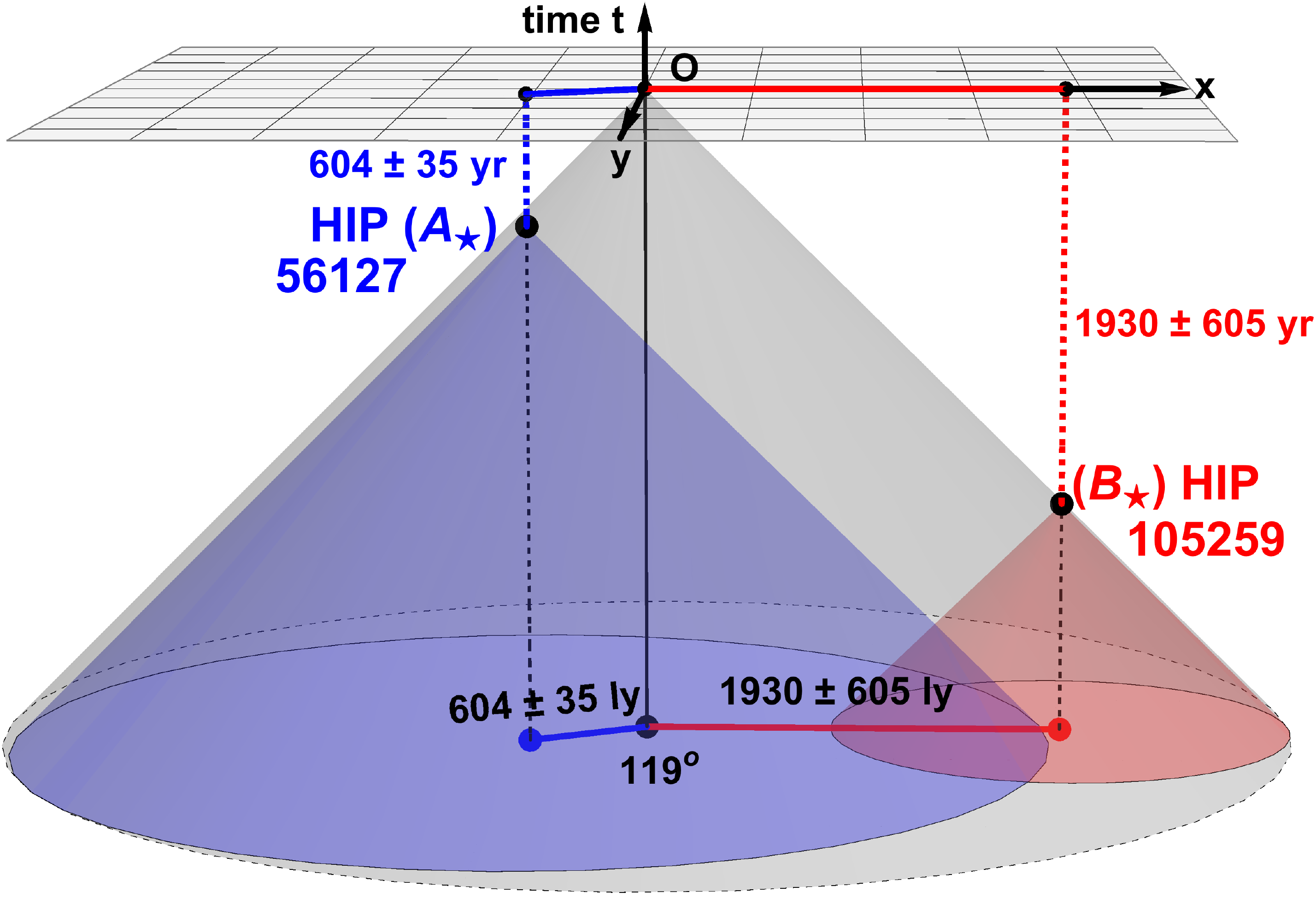}  
\end{tabular}
\vspace{-0.4cm}
\caption{
\small
2+1\,D space-time diagram for run 1 with past light cones for stellar emission events $A_{\star}$ and $B_{\star}$. Two spatial dimensions are shown ($x$-$y$ plane) with the third suppressed. The stellar pair's angular separation on the sky is the angle between the red and blue vectors. Our data rule out local-realist models with hidden variables in the gray space-time region. We do not rule out models with hidden variables in the past light cones for events $A_{\star}$ (blue), $B_{\star}$ (red), or their overlap (purple). 
\vspace{-0.4cm}
}
\label{fig:space}
\end{figure}
%%%%%%%%%%%%%%%%%%%%%%%%%%%%%%%%%%%%%%%%

%-----------------------------------------------------------------------------------------------------------------------------------------------
{\it Space-time arrangement.---} Ensuring locality requires that any information leaving Alice's star along with her setting-determining stellar photon and traveling at the speed of light could not have reached Bob before his measurement of the entangled photon is completed, and vice versa. This also achieves a necessary condition for freedom of choice.

The projected 1+1\,D space-time diagram in \fig{}~\ref{fig:space2} shows run 1 of our experiment. Entangled photons are generated at point $S$, which coincides with the origin. After local fiber transmission to the transmitting telescopes, which takes a time of $\tau_{\rm fiber}$ $\approx$180\,ns, they are sent via free-space channels to Alice at a projected distance of $x_{A} =$ 557\,m and to Bob at $x_{B} =$ 1149\,m and measured at events $A$ and $B$, respectively. Settings are determined far away at the stellar emission events $A_{\star}$ and $B_{\star}$, respectively. To close the locality loophole, entangled photon detections cannot be accepted outside a certain (maximal) time interval $\tau^{k}_{\rm valid}$ ($k=\{A,B\}$) after the detection of a stellar photon, which, in general, must be chosen such that the corresponding setting is still space-like separated from the measurement on the other side.

The time-dependent locations of the stars on the sky relative to our ground-based experimental sites make $\tau^{k}_{\rm valid}(t)$ time-dependent parameters. However, since our Rx-SPs
pointed out of windows, resulting in highly restrictive azimuth/altitude limits for the star selection, $\tau^{k}_{\rm valid}(t)$ did not change significantly during the three minutes of measurement for each experimental run. Consequently, using spatial site coordinates from Fig.~\ref{fig:sitemap} and the star's celestial coordinates (see the Supplemental Material \cite{SM-ref}), $\bar{\tau}^{k}_{\rm valid} = \min_t \{ \tau^{k}_{\rm valid}(t) \}$ was calculated as 2.55\,$\mu$s for Alice and 6.93\,$\mu$s for Bob for run 1 (see Table~\ref{tab:exp}). 

The final time-window $\tau^{k}_{\rm used}$ was chosen to be as large as possible (see Fig. 2), while also subtracting the time it takes 
to implement a setting ($\tau_{\rm set} \approx 170\,$ns) and to ensure optimal operation of the EOM (e.g.~by minimizing piezo-electric ringing). The latter required subtracting safety buffers of $\tau^A_{\rm buffer} = $ 0.38\,$\mu$s for Alice and $\tau^B_{\rm buffer} = 1.76\,\mu$s for Bob, which easily accounted for the delay of stellar photons due to the index of refraction of the atmosphere ($\tau_{\rm atm} \approx18\,$ns) \cite{stone2001index}, and any small inaccuracies in timing or the distances between the experimental stations (see the Supplemental Material \cite{SM-ref}).

In \fig{}~\ref{fig:space2}, stellar photons arriving parallel to the arrows in the blue and red shaded space-time regions (corresponding to the time intervals $\bar{\tau}^{k}_{\rm valid}$) provide valid basis settings, ensuring space-like separation for all relevant events. The darker shading corresponds to regions actually used in run 1, with $\tau^A_{\rm used} = 2\,\mu$s and $\tau^B_{\rm used} = 5\,\mu$s. In run 1, the fractions of time with valid settings for Alice and Bob were $24.9\%$ and $40.6\%$, respectively, while in run 2 they were $22.0\%$ and $44.6\%$, respectively.
Duty cycles for each observer differ primarily due to different values of $\tau^k_{\rm used}$ and different count rates for each star (see the Supplemental Material \cite{SM-ref}).

We pre-selected candidate stars within the highly restrictive azimuth and altitude limits of the stellar photon receiving telescopes from the Hipparcos catalogue \cite{perryman97,vanleeuwen07a} with parallax distances greater than $500$ ly, distance errors less than $50\%$ and Hipparcos $H_p$ magnitude between 5 and 9. Combined with the geometric configuration of the sites, selection of these stars ensured sufficient setting validity times on both sides during each experimental run of 179 seconds. To ensure a sufficiently high signal-to-noise ratio, we chose $\sim$5-$6$ magnitude stars (see the Supplemental Material \cite{SM-ref}). Note that to avoid detector saturation, parts of the entrance aperture of the Rx-SPs had to be covered. 

\fig{}~\ref{fig:space} shows a 2+1\,D space-time diagram for the stellar emission events from run 1.  Events associated with relevant hidden variables could lie within the past light cones of stellar emission events $A_{\star}$ or $B_{\star}$, the most recent of which originated $604 \pm 35$ years ago, accounting for parallax distance errors arising primarily from the angular resolution limits of the Hipparcos mission \cite{perryman97,vanleeuwen07a}.

%%%%%%%%%%%%%%%%%%%%%%%%%%%%%%%%%%%%%%%%
\linespread{1.0}
\begin{table}[ht]
\scriptsize
\centering
\begin{tabular}{| c | c | c |  r | r | r |  c | c | c | c | } 
\hline
Run & Side & HIP ID                       & az$_k^\circ$ & alt$_k^\circ$  & $d_k \!\pm\! \sigma_{d_k}\,$[ly]  & $\bar{\tau}^{k}_{\rm valid}\,$[$\mu$s]  & $\Sexp$                 & $p$-value                      & $\nu$\\
\hline
$1$ &  $A$ & 56127                  & 199 & 37  & $604 \!\pm\! 35$   & 2.55  & $2.43$         & $1.8 \!\cdot\! 10^{-13}$   & $7.3$  \\ 
    &  $B$ & 105259A  & 25 & 24  & $1930 \!\pm\! 605$   & 6.93  &  & & \\ 
 \hline
$2$ & $A$ & 80620                 & 171 & 34 & $577 \!\pm\! 40$     & 2.58   & $2.50$         & $4.0 \!\cdot\! 10^{-33}$     & $11.9$    \\
       & $B$ & 2876               & 25  &  26 & $3624 \!\pm\! 1370$ & 6.85   &  & & \\ 
\hline
\end{tabular}\par
\vspace{-0.2cm}
\caption{
\small
For Alice and Bob's side ($k=\{A,B\}$), we list Hipparcos ID numbers, Azimuth (az$_k$) (clockwise from due North) and Altitude (alt$_k$) above horizon during the observation, and parallax distances ($d_k$) with errors ($\sigma_{d_k}$) for stars observed during runs 1 and 2, which began at UTC 2016-04-21 21:23:00 and 2016-04-22 00:49:00, respectively, each lasting 179\,s. $\bar{\tau}^{k}_{\rm valid}$ is the minimum time that detector settings are valid while the star on side $k$ remained visible during each run, before subtracting delays and safety margins (see \fig{}~\ref{fig:space2}). The last 3 columns show the measured CHSH parameter for runs 1 and 2, as well as the $p$-value and the number of standard deviations $\nu$ by which our local-realist model can be rejected (see the Supplemental Material \cite{SM-ref}). 
\label{tab:exp}
}
\end{table}
%%%%%%%%%%%%%%%%%%%%%%%%%%%%%%%%%%%%%%%%

%-----------------------------------------------------------------------------------------------------------------------------------------------
{\it Analysis and results.---} We performed two cosmic Bell tests, each lasting 179 seconds. In runs 1 and 2, Alice and Bob's settings were chosen with photons from Hipparcos stars in Table~\ref{tab:exp}. To analyze the data, we make the assumptions of fair sampling and fair coincidences\ \cite{larsson14}. Thus, all data can be post-selected to include coincidence events between Alice's and Bob's measurement stations. We correct for GPS clock drift as in \cite{scheidl09} and identify coincidences within a $2.5\,$ns time window. 

\begin{figure}[t]
\begin{center}
\includegraphics[width=2.15in]{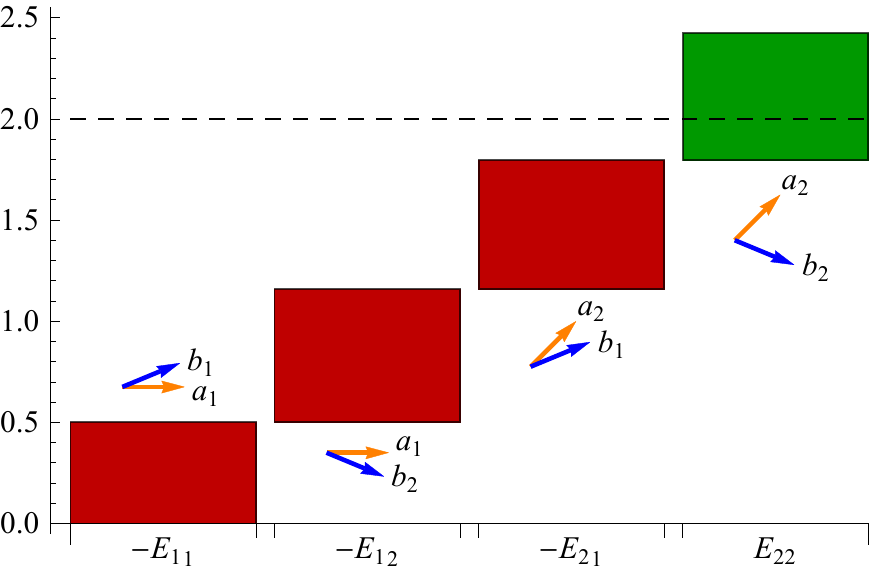}
\end{center}
\vspace{-0.7cm}
\caption{
\small
For run 1, bars shows the correlation $E_{ij}$ for each joint setting combination $(a_i,b_j)$. CHSH terms are displayed with negative signs (red) and a positive sign (green), showing that our data violate the local-realist bound (dashed line, $S=2$). The $|E_{ij}|$ values are unequal due to limited state visibility and imperfect alignment of the polarization measurement bases.
\vspace{-0.3cm}
}
\label{Fig_Histogram}
\end{figure}

We then analyze correlations between measurement outcomes $A,B \in \{+1,-1\}$ for particular setting choices $(a_i,b_j)$, $i,j \in \{1,2\}$ using the Clauser-Horne-Shimony-Holt (CHSH) inequality \cite{clauser69}: 
\begin{align}
S \equiv |E_{11}+E_{12}+E_{21}-E_{22}|\leq2, \label{eq CC}%
\end{align}
where $E_{ij}=2\,p(A\!=\!B|a_{i}b_{j})-1$ and $p(A\!=\!B|a_{i}b_{j})$ is the probability that Alice and Bob measure the same outcome given joint settings $(a_i,b_j)$. While the local-realist bound is $2$, the quantum bound is $2\sqrt{2}$, and the logical (algebraic) bound is $4$. Our run 1 data yields $\Sexp=2.425$, while for run 2, we observe $\Sexp=2.502$. Both runs therefore violate the corresponding local-realist bound. See Fig.~\ref{Fig_Histogram} and Table~\ref{tab:exp}.

Our analysis must further consider that some experimental trials will have ``corrupt" settings triggered not by genuine stellar photons, but by atmospheric airglow, thermal dark counts, errant dichroic mirror reflections, or other noise in our detectors. Since these events originate very recently in the experiments' past light cone, settings chosen with them are no better (and no worse) than settings chosen with conventional random number generators.

To constrain the fraction of experimental runs we can tolerate in which either or both sides were triggered by a corrupt event, we conservatively assume that any such events can produce maximal CHSH correlations ($S=4$) \cite{popescu94}, whereas settings triggered by correctly identified stellar photons are assumed to obey local realism ($S \leq 2$). An optimally efficient local hidden-variable model would only need to use individual corrupt photons on a single side to achieve $S=4$, without needing to ``waste" simultaneous corruptions or events in which changing the path of a stellar photon through the setting reader is unnecessary. We therefore use this maximally conservative model as our null hypothesis.

To calculate the statistical significance of our results, we account for background events and errant dichroic mirror reflections as well as differences in the measured total and noise rates for the red/blue dichroic ports on each side, which yield unequal (biased) 
frequencies for various combinations of detector settings $a_i b_j$. Moreover, whereas we assume fair sampling (for both entangled and stellar photon detections) and fair coincidences for entangled photons \cite{larsson14}, we adopt the conservative assumption that the local hidden-variable model could retain ``memory" of settings and outcomes of previous trials \cite{gill03,gill14,bierhorst15}. As detailed in the Supplemental Material \cite{SM-ref}, we find that the measured fractions of corrupt coincidences are sufficiently low that the probability that a local hidden-variable model could explain the observed violations of Bell's inequality is 
$p \leq 1.78 \times 10^{-13}$ for experimental run 1, and $p \leq 3.96 \times 10^{-33}$ for run 2. These correspond to experimental violations of the CHSH bound by at least $7.31$ and $11.93$ standard deviations, respectively.  

%%%%%%%%%%%%%%%%%%%%%%%%%%%%%%%%%%%%%%%%
%%%%%%%%%%%%%%%%%%%%%%%%%%%%%%%%%%%%%%%%

%----------------------------------------------------------------------------------------------------------------------------------------------------------------------------------
{\it Conclusions.---} For both runs, we assume fair sampling, close the locality loophole and, for the first time, explicitly constrain freedom of choice with astronomically chosen settings, relegating any local-realist models to have acted no more recently than $604 \pm 35$ and $577 \pm 40$ years ago, for runs 1 and 2, respectively. Therefore, any hidden-variable mechanism exploiting the freedom-of-choice loophole would need to have been enacted prior to Gutenberg's invention of the printing press, which itself pre-dates the publication of Newton's {\it Principia} by two and a half centuries. While a Bell test like ours only constrains a local-realist mechanism to have acted no later than the most recent of the astronomical emission events, we note that any process that requires both emission events to have been influenced by the same common cause would be relegated to even earlier times, to when the past light cones from each emission event intersected $2409 \pm 598$ and $4040 \pm 1363$ years ago, for runs 1 and 2, respectively \cite{friedman13a}. 

This work thus represents the first experiment to dramatically limit the space-time region in which hidden variables could be relevant, paving the way for future ground- and space-based tests with distant galaxies, quasars, patches of the cosmic microwave background (CMB), or other more exotic sources such as neutrinos and gravitational waves. Such tests could progressively push any viable hidden-variable models further back into deep cosmic history \cite{gallicchio14}, billions of years in the case of quasars, back to the early universe in the case of the CMB, or even, in the case of primordial gravitational waves, further back to any period of inflation preceding the conventional big bang model \cite{guth81,linde82,albrecht82,guth05,guth14}. 

%----------------------------------------------------------------------------------------------------------------------------------------------------------------------------------
{\it Acknowledgements.---}
The authors would like to thank Brian Keating, Michael J.W. Hall, Jan-{\AA}ke Larsson, Marissa Giustina, Ned Hall, Craig Callender, Jacob Barandes, David Kagan, and Larry Guth for useful discussions. Thanks to the Austrian National Bank (OENB) and the Bundesimmobiliengesellschaft (BIG) for providing the rooms for our receiving stations. This work was supported by the Austrian Academy of Sciences (OEAW), by the Austrian Science Fund (FWF) with SFB F40 (FOQUS) and FWF project CoQuS No.~W1210-N16 and the Austrian Federal Ministry of Science, Research and Economy (BMWFW).  A.S.F, D.I.K, A.H.G, and J.G. acknowledge support for this project from NSF INSPIRE Award PHY-1541160. A.S.F. acknowledges support from NSF Award SES-1056580. A.M. and I.S. acknowledge support from MIT's Undergraduate Research Opportunities Program (UROP). Portions of this work were conducted in MIT's Center for Theoretical Physics and supported in part by the U.S. Department of Energy under grant Contract Number DE-SC0012567. J.G. acknowledges support from NSF Award PLR-1248097 and Harvey Mudd College.

\clearpage

%----------------------------------------------------------------------------------------------------------------------------------------------------------------------------------
\onecolumngrid
\appendix

\centerline{\large\bf Supplemental Material} 
\vspace{0.15cm}
\centerline{\large\bf Cosmic Bell Test: Measurement Settings from Milky Way Stars}
\vspace{0.3cm}

\twocolumngrid

%-----------------------------------------------------------------------------------------------------------------------------------------------
\vspace{\skips}
\section{Causal Alignment}
\label{sec:caus1}
\vspace{\skips}
%-----------------------------------------------------------------------------------------------------------------------------------------------

Compared to a standard Bell test, the time-dependent locations of the stars on the sky relative to our ground-based experimental sites complicate the enforcement of the space-like separation conditions needed to address both the locality and freedom-of-choice loopholes. For example, the photon from star $A_\star$ must be received by Alice's stellar photon receiving telescope (Rx-SP) before that photon's causal wave front reaches either the Rx-SP or the entangled photon receiving telescope (Rx-EP) on Bob's side, and vice versa. 

To compute the time-dependent durations $\tau^{k}_{\rm valid} (t)$ (for $k = \{A, B\}$) that settings chosen by astronomical photons remain valid, we adopt a coordinate system with the center of the Earth as the origin. The validity times on each side due 
to the geometric configuration of the stars and ground-based sites are then given by
\begin{eqnarray}
\tau_{\rm valid}^{A}(t) & =& \frac{1}{c}\qaa(t) \cdot (\raa - \mbb) + \frac{\nair}{c} \Big[ | \maa - \sepr | - | \mbb - \sepr | \Big] \nonumber \\ 
              &\quad&  -  \frac{\na}{c} |\raa - \maa|  \nonumber \\
\tau_{\rm valid}^{B}(t) & =&  \frac{1}{c}\qbb(t) \cdot (\rbb - \maa) + \frac{\nair}{c} \Big[ | \mbb - \sepr | - | \maa - \sepr | \Big] \nonumber \\ 
              &\quad&  - \frac{\nb}{c} |\rbb - \mbb| ,
\label{Deltas_diff2}
\end{eqnarray}
where $\raa$ is the spatial 3-vector for the location of Alice's Rx-SP, $\maa$ is the spatial 3-vector for Alice's Rx-EP (and likewise for $\rbb$ and $\mbb$ on Bob's side), $\sepr$ is the location of the entangled photon source, and $c$ is the speed of light in vacuum. The time-dependent unit vectors $\qaa (t) , \qbb (t)$ point toward the relevant stars, and are computed using astronomical ephemeris calculations. Additionally, $\nair$ is the index of refraction of air and $\na,\nb$ parametrize the group velocity delay through fiber optics / electrical cables connecting the telescope and entangled photon detectors. 
To compute $\tau_{\rm valid}^{k} (t)$, we make the reasonable approximation that the Rx-SP and Rx-EP are at the same spatial location on each side, such that $\raa=\maa$ and $\rbb=\mbb$, and the computations require the GPS coordinates of only 3 input sites (see Table~\ref{tab:sites}). This assumes negligible delays from fiber and electrical cables via the $\na,\nb$ terms.
Negative validity times $\tau^{k}_{\rm valid}(t)$ for either side would indicate an instantaneous configuration that was out of ``causal alignment," in which at least one setting would be invalid for the purposes of closing the locality loophole. For runs 1 and 2, $\tau^{k}_{\rm valid}(t) > 0$ for the entire duration of 179\,s, with minimum times in Table I.

We subtract the time it takes to implement a setting with the electro-optical modulator, $\tau_{\rm set} \approx$ 170\,ns, and subtract additional conservative buffer margins $\tau^{k}_{\rm buffer}$ (0.38\,$\mu$s for Alice and $1.76$\,$\mu$s for Bob) to determine the minimum time windows $\tau^{k}_{\rm used}$ in Eq.~(\ref{tau_used}) utilized during the experiment (see Table~\ref{tab:exp1}):
\begin{eqnarray}
%\tau^{k}_{\rm used} & =&  \min_t \Big\{ \tau^{k}_{\rm valid} (t)\Big\} - \tau_{\rm set} - \tau_{\rm atm} - \tau^{k}_{\rm buffer},
\tau^{k}_{\rm used} & =&  \min_t \Big\{ \tau^{k}_{\rm valid} (t)\Big\} - \tau^{k}_{\rm buffer} - \tau_{\rm set},
\label{tau_used}
\end{eqnarray}
where $\tau_{\rm set}$ includes the total delays on either side due to reflections inside the telescope optics, the SPAD detector response, and electronic readout on the astronomical receiver telescope side as well as the time to switch the Pockels cell and electronically use the FPGA board to output a random number. The next section conservatively estimates $\tau_{\rm atm}$ $\approx 18$\,ns for the delay due to the index of refraction of the atmosphere for either observer.
While $\tau_{\rm atm}$ is not explicitly considered in Eq.~(\ref{tau_used}), 
it is well within
the buffer margins, since $\tau^{k}_{\rm buffer} \gg \tau_{\rm atm}$, which also encompass any small inaccuracies in the timing or distances between the experimental sites.

Although $\tau_{\rm valid}^k (t)$ depends on time, motivating our use of $\bar{\tau}_{\rm valid}^k \equiv {\rm min}_t \{ \tau_{\rm valid}^k (t) \}$ when computing $\tau_{\rm used}^k$, the actual values of $\tau_{\rm valid}^k$ changed very little during our observing windows. For the stars used in experimental run 1, $\Delta \tau_{\rm valid}^A = 2.96$ ns and $\Delta \tau_{\rm valid}^B = 17.26$ ns; for experimental run 2, $\Delta \tau_{\rm valid}^A = 18.97$ ns and $\Delta \tau_{\rm valid}^B = 17.27$ ns. The largest of these differences represents less than $1\%$ of the relevant $\bar{\tau}_{\rm valid}^k$.

By ensuring that the Pockels cell switched if it had not been triggered by a fresh setting within the last 
$\tau^A_{\rm used} = 2$\,$\mu$s for Alice and $\tau^B_{\rm used} = 5$\,$\mu$s for Bob,
we only record and analyze coincidence detections for entangled photons obtained while the settings on both sides remain valid.

%%%%%%%%%%%%%%%%%%%%%%%%%%%%%%%%%%%%%%%%
\begin{figure*}
\centering
\begin{tabular}{@{}c@{}c@{}c@{}}
\includegraphics[width=2.2in]{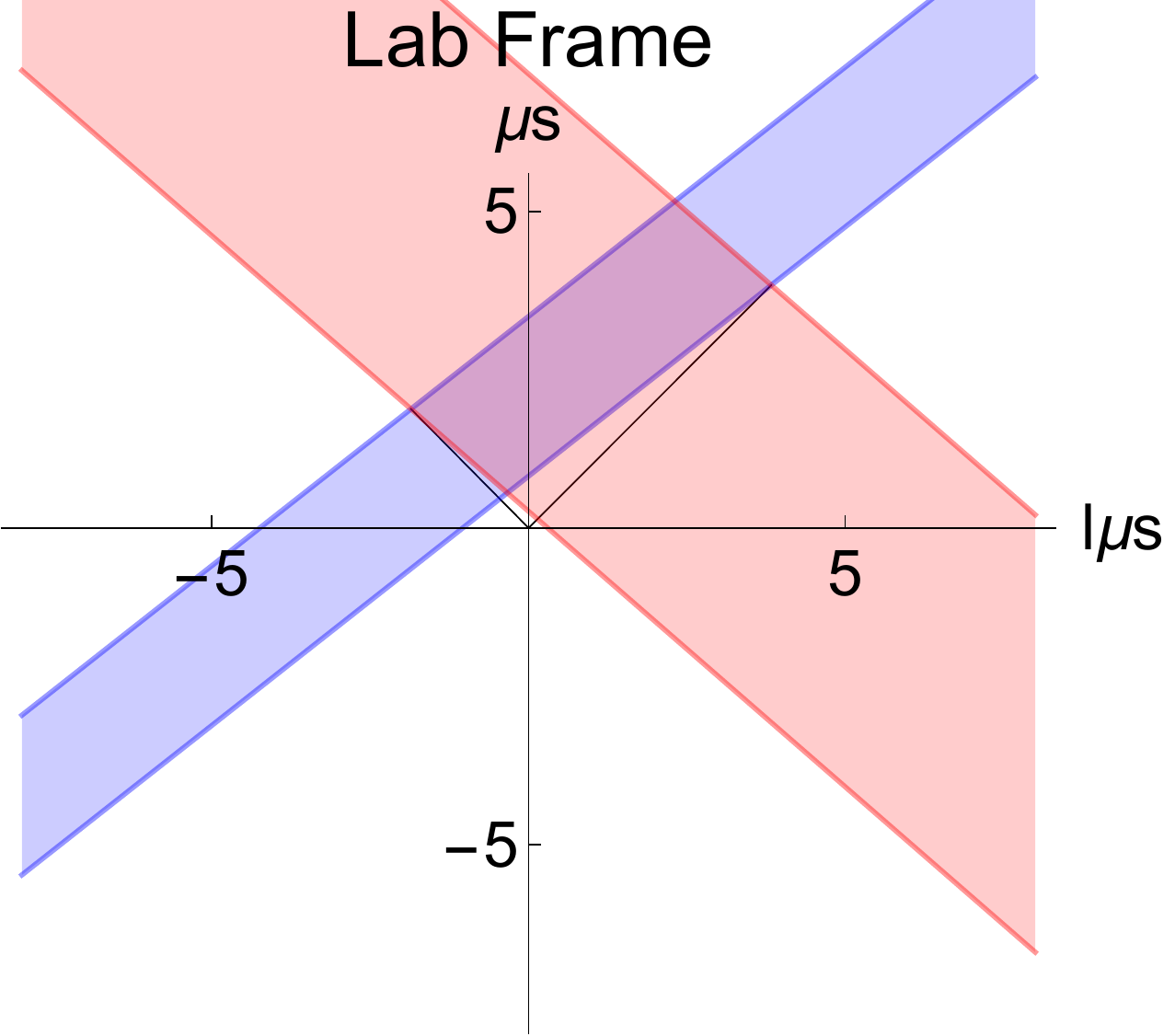}  &
\includegraphics[width=2.2in]{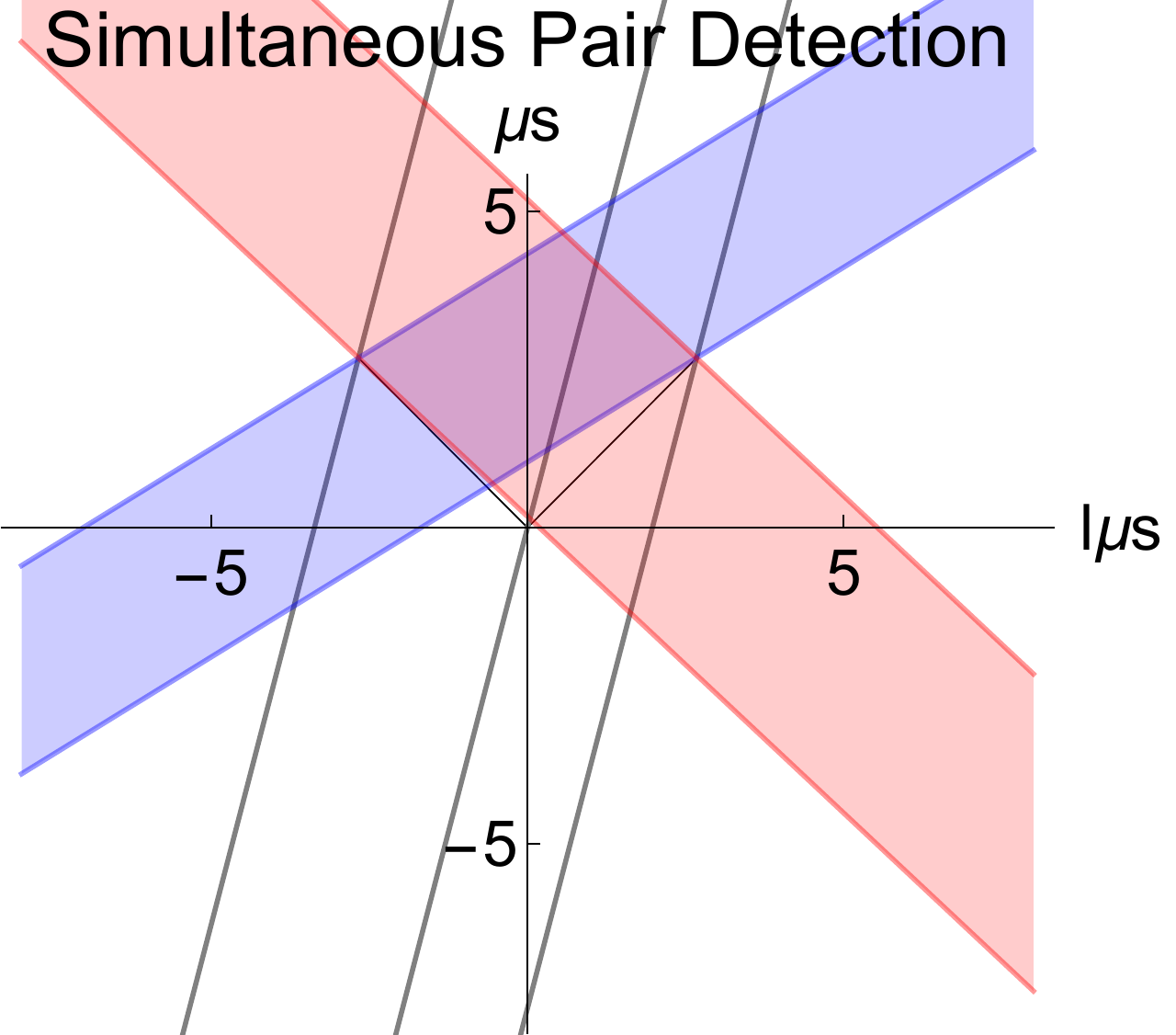}  &
\includegraphics[width=2.2in]{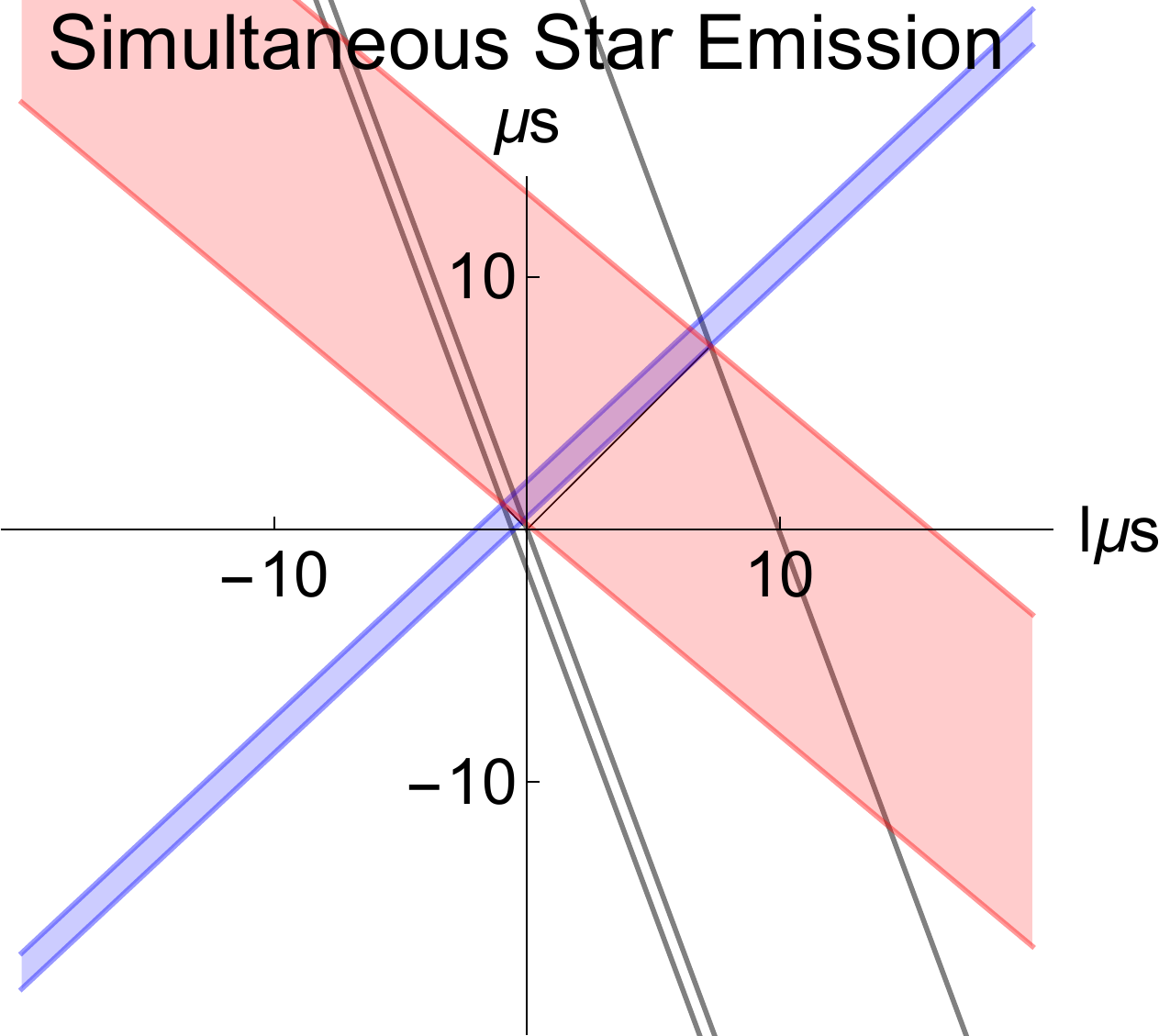}  \\
\includegraphics[width=2.2in]{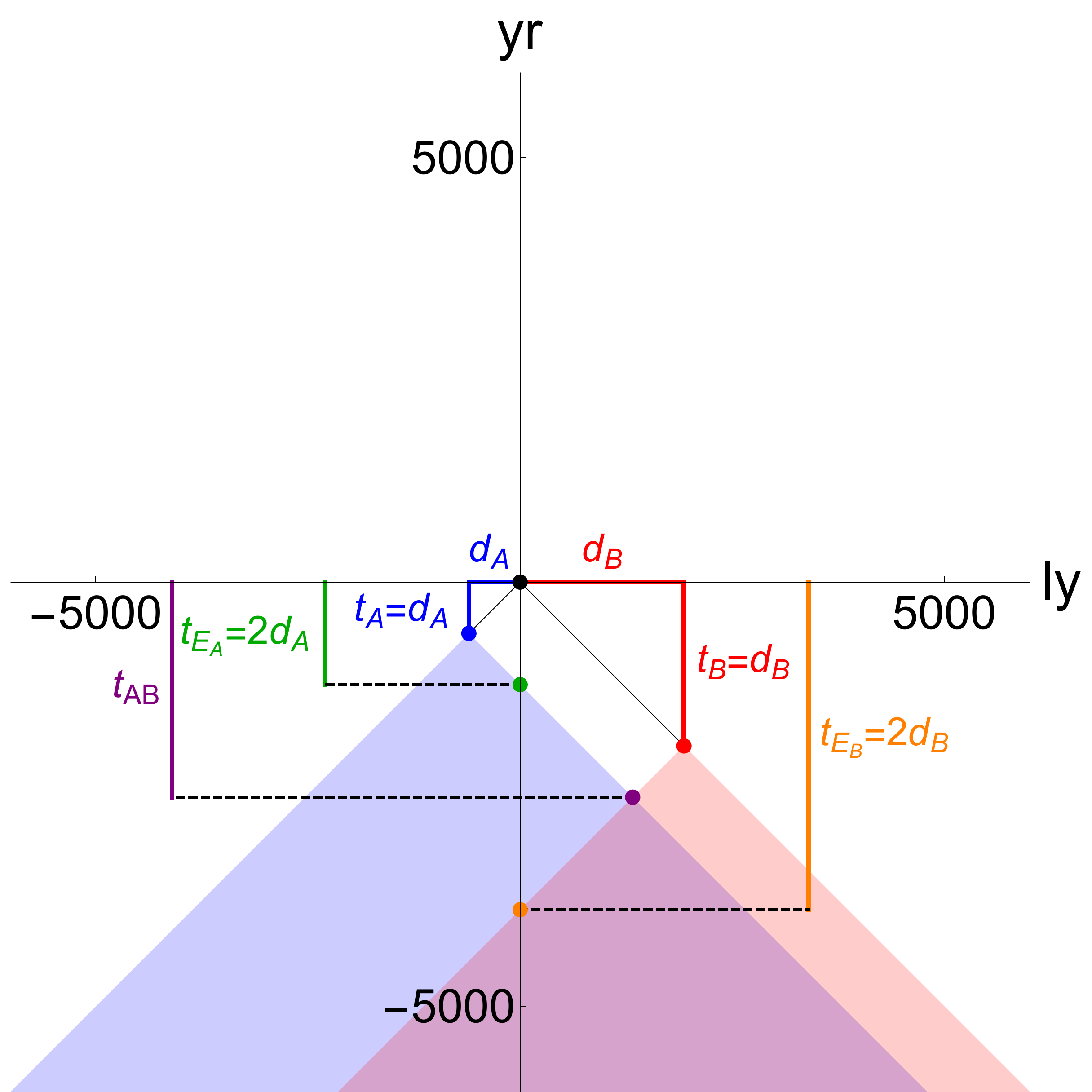}  &
\includegraphics[width=2.2in]{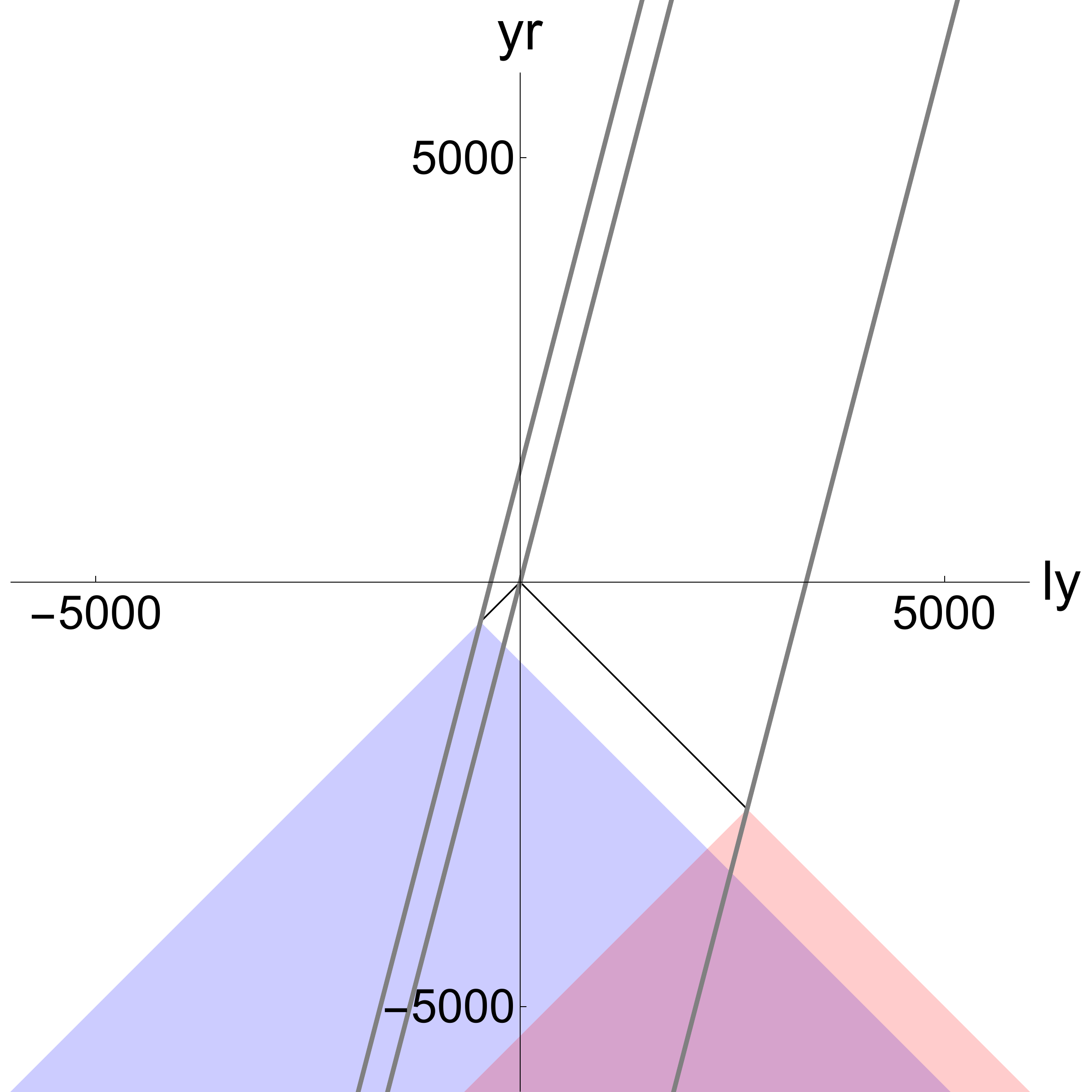}  &
\includegraphics[width=2.2in]{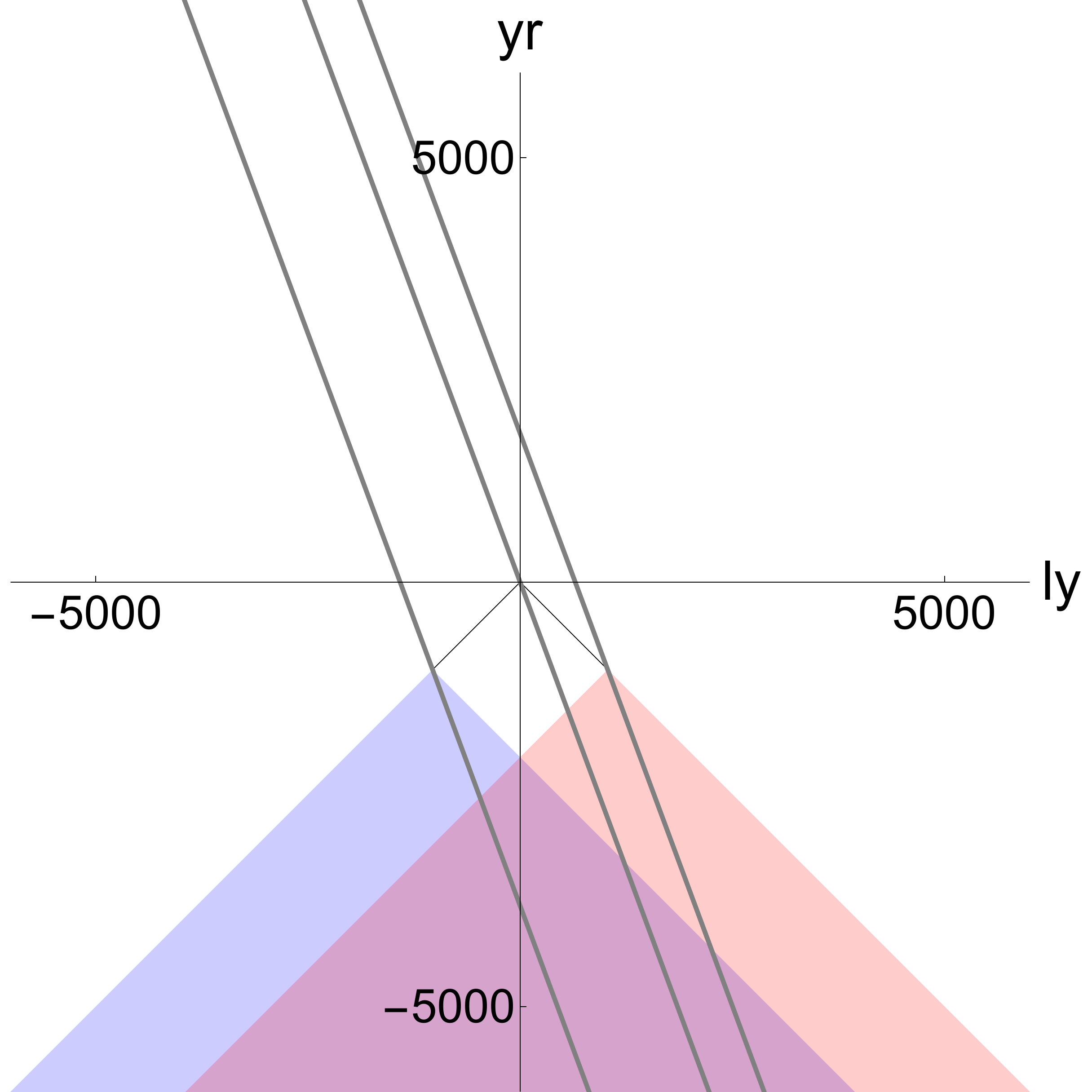}  \\
\end{tabular}
\vspace{-0.4cm}
\caption{
\small
Run 1's 1+1\,D space-time diagrams (with time on the y-axis and one spatial dimension on the x-axis) are shown in each column, left to right, for 3 frames: (1) laboratory, (2) simultaneous entangled photon pair detection, and (3) simultaneous stellar photon emission. Relevant space-time regions are shown for Alice (blue) and Bob (red). The spatial axis onto which all events were projected (red line in \fig{}~\ref{fig:map} for lab frame) was chosen to minimize its distance to Alice and Bob. Slanted black lines in the second and third columns indicate Lorentz boosts relative to the lab frame in the first column. ({\it Top row}) Space-time diagrams for the experiment. Solid blue and red areas denote space-time regions for each stellar photon to provide a valid detector setting in each frame. Scales for the x- and y-axes are in units of light-microseconds and microseconds, respectively. ({\it Bottom row}) Zoomed out space-time diagrams with past light cones for the stellar emission events observed by Alice (blue) and Bob (red). Units for the x- and y-axes are in light years and years, respectively. The lower left panel includes labels for quantities computed in the Lookback Times secion of the SM using the {\it projected} stellar distances along the chosen spatial axis (with projected angular separation $\alpha=180^{\circ}$). The upper diagrams zoom in at the tip of the light cone at the origin of the bottom row plots. 
}
\label{fig:space3}
\end{figure*}
%%%%%%%%%%%%%%%%%%%%%%%%%%%%%%%%%%%%%%%%

%----------------------------------------------------------------------------------------------------------------------------------------------------------------------------------
\subsubsection{Atmospheric Delay}
\label{sec:delay}
\vspace{\vskips}
%----------------------------------------------------------------------------------------------------------------------------------------------------------------------------------
The air in the atmosphere causes a relative delay between the causal light cone, which expands outward at speed $c$, and the photon, which travels at $c/n$, where $n$ is the index of refraction of air. We estimate this effect by computing the light's travel time through the atmosphere on the way to the observer. If the atmosphere has index of refraction $n$ and scale height $z_0$, the delay time is
\begin{eqnarray}
\Delta t = \frac{(z_0 - h) X (n-1)}{c}
\label{eq:dt}
\end{eqnarray}
where $X$ is the airmass. The minimum elevation of each stellar photon receiving telescope is $h = \SI{200}{m}$ above sea level, and the minimum altitude angle is $\delta = 24^\circ$ above the horizon with airmass $X\approx 2.5$ (see Tables~\ref{tab:sites}-\ref{tab:exp1}). Neglecting the earth's curvature (which is a conservative approximation), we use $z_0 =$ 8.0\,km and the index of refraction at sea level of $n-1 \approx 2.7\times 10^{-4}$~\cite{stone2001index}. The delay between the arrivals of the causal light cone and the photon itself may be conservatively estimated to be $\Delta t=17.6$\,ns due to the atmosphere. 
%----------------------------------------------------------------------------------------------------------------------------------------------------------------------------------

%----------------------------------------------------------------------------------------------------------------------------------------------------------------------------------
\vspace{\skips}
\subsection{Source Selection}
\label{sec:stars}
\vspace{\skips}
%----------------------------------------------------------------------------------------------------------------------------------------------------------------------------------

We used custom Python software to select candidate stars from the Hipparcos catalogue \cite{perryman97,vanleeuwen07a} with parallax distances greater than $500$\,ly, distance errors less than $50\%$, and Hipparcos $H_p$ magnitude between 5 and 9 to avoid detector saturation and ensure sufficient detection rates. Telescopes pointed out of open windows at both sites (see Table~\ref{tab:sites}). A list of $\sim$$100$-$200$ candidate stars were pre-selected per side for each night due to the highly restrictive azimuth/altitude limits.  Candidate stars were visible through the open windows for $\sim 20$-$50$ minutes on each side. 

Due to weather, seeing conditions, and the uncertainties in aligning the transmitting and receiving telescope optics for the entangled photon source, it was not possible to pre-select specific star pairs for each experimental run at a predetermined time. Instead, when conditions were stable, we selected the best star pairs from our pre-computed candidate lists that were currently visible through both open windows, ranking stars based on brightness, distance, the amount of time each would remain visible, the settings validity time, and the airmass at the time of observation. The 4 bright stars we actually observed for runs 1 and 2 were $\sim$5-$6$ mag (see Table~\ref{tab:exp1}). Combined with the geometric configuration of the sites (see Table~\ref{tab:sites}), selection of these stars ensured sufficient setting validity times on both sides during each experimental run of 179 seconds. 

%%%%%%%%%%%%%%%%%%%%%%%%%%%%%%%%%%%%%%%%
\linespread{1.0}
\begin{table}[ht]
\footnotesize
\centering
\begin{tabular}{| c | c | c | c | c | } 
\hline
Site & Lat.$^\circ$                     & Lon.$^\circ$ & Elev. [m]   & Telescope [m]  \\
\hline
Telescope $A$ &  48.21645                & 16.354311 & 215.0     &  0.2032        \\
Source $S$ & 48.221311               & 16.356439 & 205.0    & \ldots       \\
Telescope $B$ &  48.23160                 & 16.3579553 & 200.0   & 0.254        \\
\hline
\end{tabular}\par
\vspace{-0.2cm}
\caption{
\small
Latitude, Longitude, Elevation, for Alice ({\it A}), Bob ({\it B}) and the Source ({\it S}), and aperture diameter of the stellar photon receiving telescopes. 
\label{tab:sites}
\vspace{-0.3cm}
}
\end{table}
%%%%%%%%%%%%%%%%%%%%%%%%%%%%%%%%%%%%%%%%

%%%%%%%%%%%%%%%%%%%%%%%%%%%%%%%%%%%%%%%%
%TABLE 
%%%%%%%%%%%%%%%%%%%%%%%%%%%%%%%%%%%%%%%%
\linespread{1.0}
\begin{table*}[ht]
\footnotesize
\centering
\begin{tabular}{| c | c  | c | r | r | c | r | r | r |  c | c | c | c | c | } 
\hline
Run & Side & HIP ID                      & RA$^\circ$  & DEC$^\circ$  & $H_p$ & az$_k^\circ$ & alt$_k^\circ$  & $d_k \pm \sigma_{d_k}$ \ [ly] & $\bar{\tau}^k_{\rm valid}$ [$\mu$s]  & Trials      & $\Sexp$                 & $p$-value                      & $\nu$\\
\hline           
$1$ &  $A$ & 56127                 & 172.5787 & -3.0035  & 4.8877 & 199  & 37  & $604 \pm 35$      & 2.55      & 136\,332 & $2.425$         & $1.78 \times 10^{-13}$   & $7.31$  \\           
    & $B$ & 105259A  & 319.8154 & 58.6235  & 5.6430 & 25   & 24  & $1930 \pm 605$    & 6.93      &         &                 &                                         &  \\ 
 \hline
$2$ & $A$ & 80620                  & 246.9311 & -7.5976  &  5.2899 & 171 & 34  & $577 \pm 40$      & 2.58      & 88\,779 & $2.502$         & $3.96 \times 10^{-33}$  & $11.93$   \\
    & $B$ & 2876                   & 9.1139   & 60.3262  & 5.8676  & 25  &  26 & $3624 \pm 1370$   & 6.85      &         &                 &                                         & \\ 
\hline
\end{tabular}\par
\vspace{-0.2cm}
\caption{
\small
More complete version of main text Table I.
For Alice and Bob's side, we list Hipparcos ID numbers, celestial coordinates, Hipparcos $H_p$ band magnitude, Azimuth (clockwise from due North) and Altitude above horizon during the observation, and parallax distances ($d_k$) with errors ($\sigma_{d_k}$) for stars observed during runs 1 and 2, which began at UTC 2016-04-21 21:23:00 and 2016-04-22 00:49:00, respectively, each lasting 179\,s.  $\bar{\tau}^k_{\rm valid}=\min_t \{ \tau^{k}_{\rm valid} (t) \}$ from Eq.~(\ref{Deltas_diff2}) is the minimum time that detector settings are valid 
for side $k=\{A,B\}$ 
during each experimental run, before subtracting delays and safety margins
(see Eqs.~(\ref{Deltas_diff2})-(\ref{tau_used}) and main text \fig{}~2). 
Star pairs for runs 1 and 2 have angular separations $\alpha$ of $119^\circ$ and $112^\circ$ on the sky, with past light cone intersection events occurring $2409 \pm 598$ and $4040 \pm 1363$ years ago, respectively. The last 4 columns show the number of double coincidence trials, the measured CHSH parameter $\Sexp$, as well as the $p$-value and number of standard deviations $\nu$ by which the null hypothesis may be rejected, based on the Method 3 analysis, below.
\label{tab:exp1}
}
\end{table*}

%%%%%%%%%%%%%%%%%%%%%%%%%%%%%%%%%%%%%%%%
\begin{figure*}
\centering
\begin{tabular}{@{}c@{}c@{}}
\includegraphics[width=3.1in]{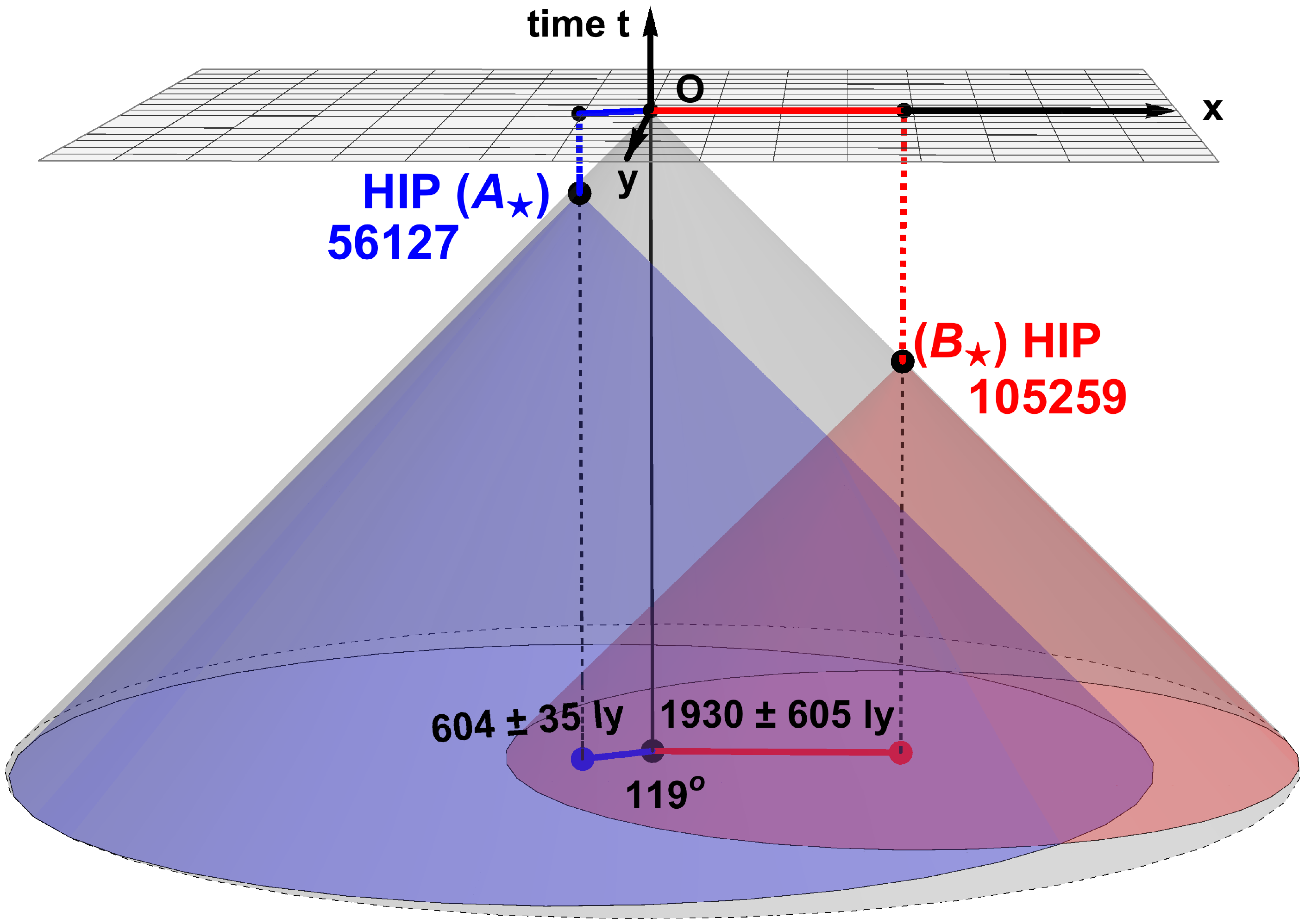}  &
\includegraphics[width=3.15in]{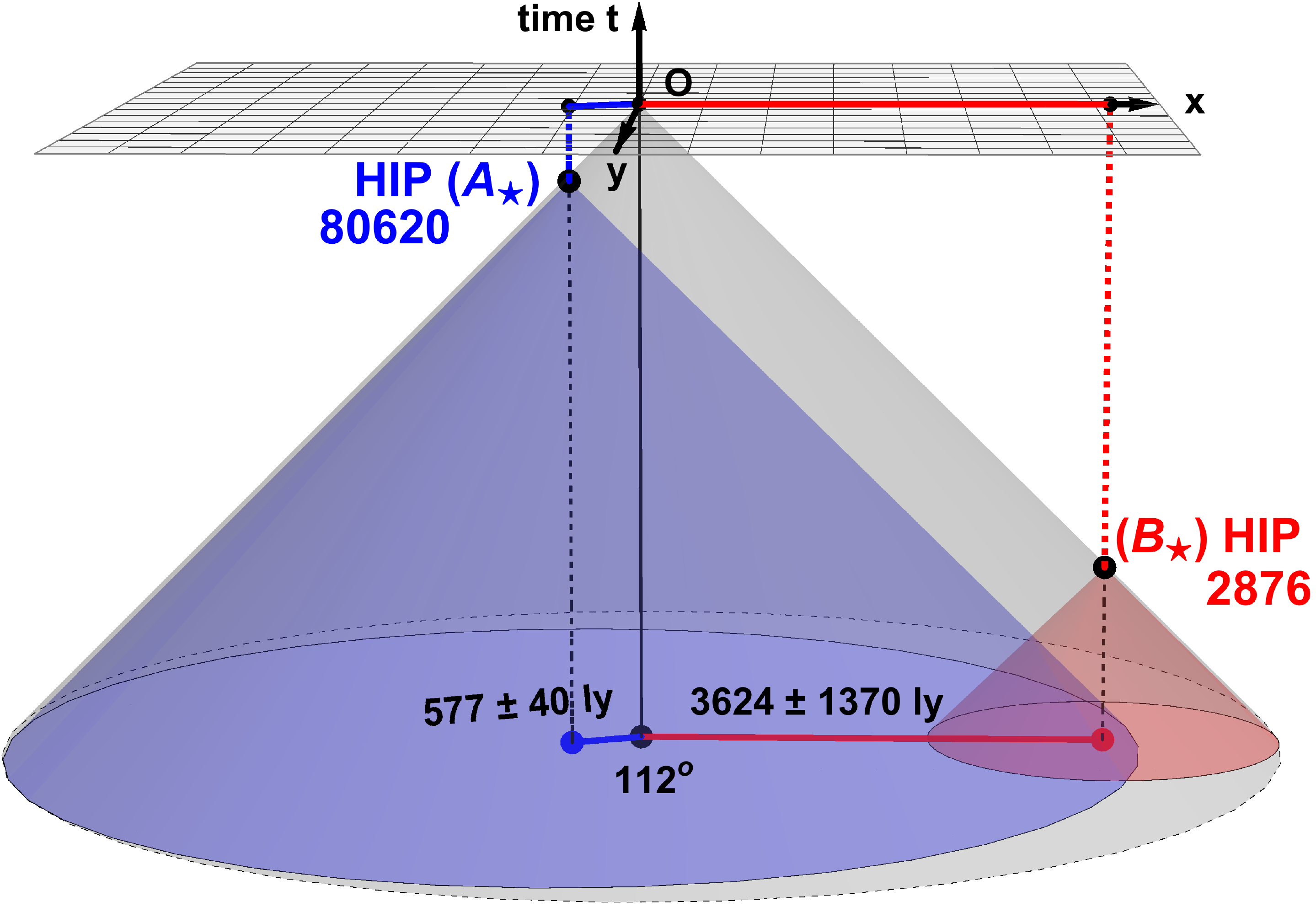}  \\
\end{tabular}
\vspace{-0.4cm}
\caption{
\small
2+1\,D space-time diagrams with past light cones for stellar emission events $A_{\star}$ and $B_{\star}$ for experimental run 1 ({\it left}) and run 2 ({\it right}). The time axis begins 5000 years before event $O$ on Earth today, with 2 spatial dimensions in the $x$-$y$ plane and the third suppressed. The stellar pair's angular separation on the sky is the angle between the red and blue vectors. Our data rule out local-realist models with hidden variables in the gray space-time regions. We do not rule out models with hidden variables in the past light cones for events $A_{\star}$ (blue), $B_{\star}$ (red), or their overlap (purple). Given the earlier emission time $B_\star$ for run 2, that run excludes models with hidden variables in a larger space-time region than run 1 (modulo the parallax distance errors in Table~\ref{tab:exp1}). 
}
\label{fig:space4}
\end{figure*}
%%%%%%%%%%%%%%%%%%%%%%%%%%%%%%%%%%%%%%%%

%%%%%%%%%%%%%%%%%%%%%%%%%%%%%%%%%%%%%%%%
\begin{figure}
\centering
\begin{tabular}{@{}c@{}}
\includegraphics[width=3.1in]{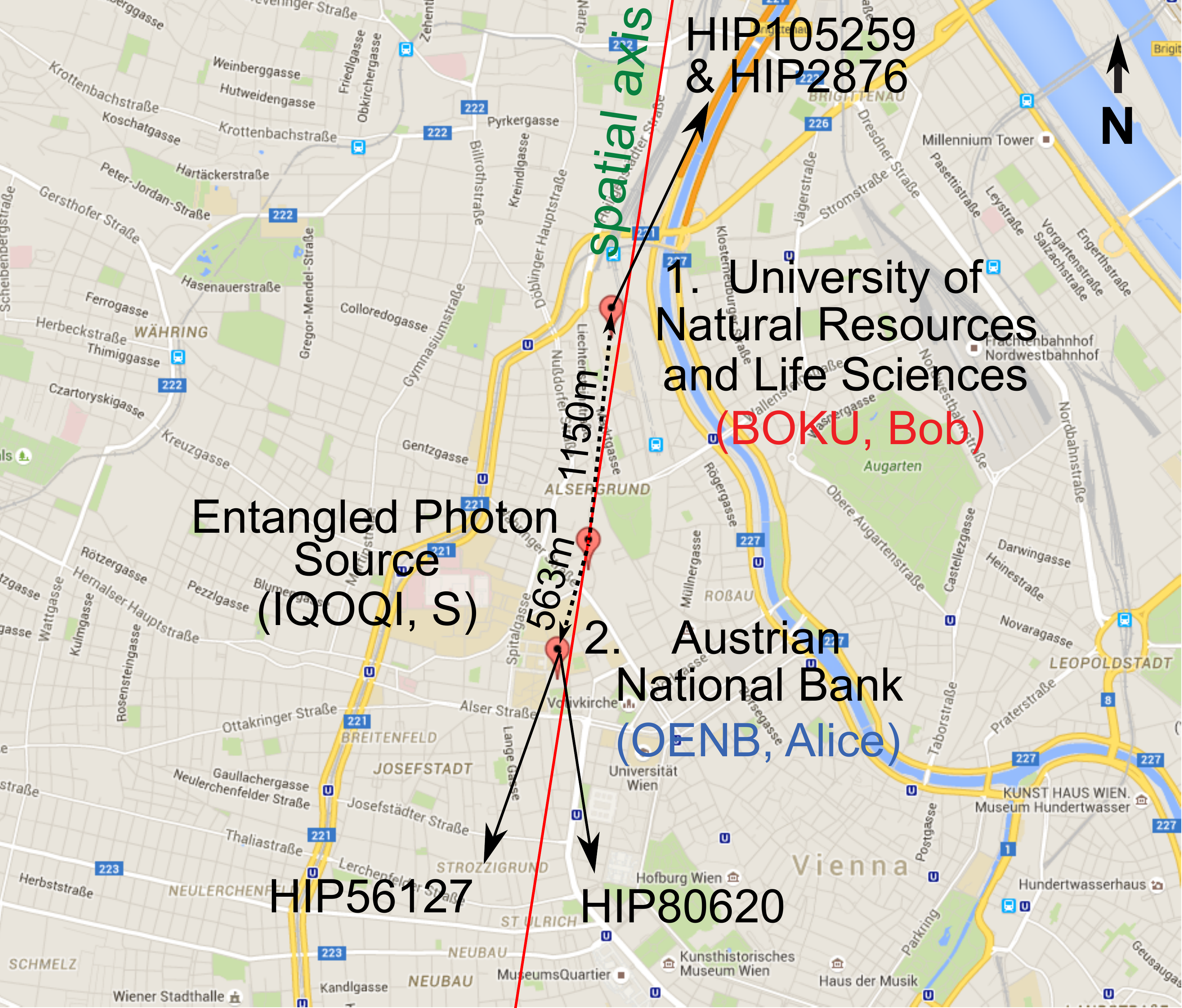}  \\
\end{tabular}
\vspace{-0.4cm}
\caption{
\small
Overhead view of Vienna with experimental sites from Table~\ref{tab:sites}.  Azimuthal directions of stars observed during runs 1 and 2 are shown (see Table~\ref{tab:exp1}). The red line denotes the projected spatial axis for the $1+1$\,D space-time diagrams in Figure
%s~\ref{fig:space2} and
~\ref{fig:space3}. (Background graphic taken from Google Maps, 2016.)
\vspace{-0.5cm}
}
\label{fig:map}
\end{figure}
%%%%%%%%%%%%%%%%%%%%%%%%%%%%%%%%%%%%%%%%

%----------------------------------------------------------------------------------------------------------------------------------------------------------------------------------
\vspace{\skips}
\subsection{Lookback Times}
\label{sec:dist_t_z_euclidean}
\vspace{\skips}
%----------------------------------------------------------------------------------------------------------------------------------------------------------------------------------

For stars within our own galaxy, the lookback time $t_k$ to a stellar emission event from a star $d_k$ light years away is $t_k=d_k$ years.  For example, Hipparcos Star HIP 2876 is located $d_B=3624$ light years (ly) from Earth, and its photons were therefore emitted $t_B=3624$ years prior to us observing them (see Table~\ref{tab:exp1}). The lookback time $t_{E_{k}}$ to when the past light cone of a stellar emission event from star $k$ intersects Earth's worldline is $t_{E_{k}} = 2 d_k$ years.

The lookback time to the past light cone intersection event $t_{AB}$ (in years) for a pair of 
Hipparcos stars is \cite{friedman13a}
\begin{equation}
t_{AB} =  \frac{1}{2} \left( d_A + d_B + \sqrt{ d_A^2 + d_B^2 - 2 d_A d_B \cos(\alpha) } \right),
\label{eq:t_l_ab_stars1}
\end{equation}
where $d_A,d_B$ are the distances to the stars (in ly) and $\alpha$ is the angular separation (in radians) of the stars, as seen from Earth. See the lower left panel of Fig.~\ref{fig:space3}.

Ignoring any covariance between $d_A$, $d_B$, and $\alpha$, and assuming the error on $\alpha$ ($\sigma_{\alpha}$) is negligible compared to the distance errors ($\sigma_{d_i}$), the 1$\sigma$ lookback time error is approximately given by
\begin{eqnarray}
\sigma_{t_{AB}} & \approx & \frac{ \sqrt{   \sum_{(i,j)}  \sigma^2_{d_i} \left[t_{AB} - \frac{d_j}{2}(1+\cos \alpha)\right]^2 } }{ 2 t_{AB} - d_A - d_B } ,
\label{eq:t_l_ab_err}
\end{eqnarray}
where $(i,j) \in \{(A,B),(B,A)\}$.

%----------------------------------------------------------------------------------------------------------------------------------------------------------------------------------
\vspace{\skips}
\subsection{Experimental Details}
\label{sec:exp2}
\vspace{\skips}
%----------------------------------------------------------------------------------------------------------------------------------------------------------------------------------

The entangled photon source was based on type--II spontaneous parametric down conversion (SPDC) in a periodically poled KTiOPO$_4$ (ppKTP) crystal with $25$\,mm length. Using a laser at $405$\,nm, the ppKTP crystal was bi-directionally pumped inside a polarization Sagnac interferometer generating degenerate polarization entangled photon pairs at $810$\,nm. We checked the performance of the SPDC source via local measurements at the beginning of each observation night. Singles and coincidence rates of approximately $1.1$\,MHz and $275$\,kHz, respectively, correspond to a local coupling efficiency (i.e., coincidence rate divided by singles rate) of roughly $25$\%. In run 1 (run 2), the duty cycle of Alice's and Bob's measurements -- i.e., the temporal sum of used valid setting intervals divided by the total measurement time per run -- were $24.9\%$ ($22.0\%$) and $40.6\%$ ($44.6\%$), respectively, resulting in a duty-cycle for valid coincidence detections between Alice and Bob of $10.1\%$ ($9.8\%$). From the measured 136\,332 (88\,779)
total valid coincidence detections per run, we can thus infer the total two-photon attenuation through the quantum channels to Alice and Bob of 15.3\,dB (16.8\,dB).

%----------------------------------------------------------------------------------------------------------------------------------------------------------------------------------
\vspace{\skips}
\section{Quality of Setting Reader}
\label{sec:spectra}
\vspace{\skips}
%----------------------------------------------------------------------------------------------------------------------------------------------------------------------------------

The value of the observed CHSH violation is highly sensitive to the fraction of generated settings which were in principle ``predictable'' by a local hidden-variable model. For this reason, it is important to have a high-fidelity spectral model of the setting generation process. In our analysis, we conservatively assume that local noise and  incorrectly generated settings are completely predictable and exploitable. An incorrectly generated setting is a red photon that generates a blue setting (or vice versa) by ending up at the wrong SPAD.

In this section we compute the fractions of incorrectly generated settings $f_{r \to b}$ and $f_{b \to r}$. For example, $f_{r\rightarrow b}$ is the conditional probability that a red photon goes the wrong way in the dichroic and ends up detected as a blue photon, generating the wrong setting. These fractions are highly sensitive to the transmission and reflection spectra of the two dichroic mirrors in each setting generator. They are somewhat less dependent on the spectral distribution of photons emitted by the 
astronomical source, on absorption and scattering in the Earth's atmosphere, the anti-reflection coatings on the optics, and the SPAD quantum efficiencies. 

A system of dichroic beamsplitters which generates measurement settings from photon wavelengths can be modeled by two functions $\rho_{\rm red}(\lambda)$ and $\rho_{\rm blue}(\lambda)$, the probability of transmission to the red and blue arms as a function of photon wavelength $\lambda$. Ideally, photons with wavelength $\lambda$ longer than some cutoff $\lambda'$ would not arrive at the blue arm: $\rho_{\rm blue}(\lambda) = 0$ for $\lambda > \lambda'$. Similarly, $\rho_{\rm red}(\lambda) = 0$ for $\lambda \leq \lambda'$ would ensure that blue photons do not arrive at the red arm. Due to imperfect dichroic beamsplitters, however, it is impossible to achieve $\rho_{\rm blue}(\lambda >\lambda') = 0$ and $\rho_{\rm red}(\lambda \leq \lambda') = 0$. 

The total number of blue settings generated by errant red photons can be computed as
\begin{equation}
N_{r\to b}(\lambda') = \int_{\lambda'}^{\infty} \rho_{\text{blue}}(\lambda) \> N_{\text{in}}(\lambda)~d\lambda ,
\end{equation}
where $N_{\text{in}}(\lambda)$ is the spectral distribution of the stellar photons remaining after losses due to the atmosphere, anti-reflection coatings, and detector quantum efficiency. Then the fraction $f_{r\to b}$ can be computed by normalizing 
\begin{equation}
f_{r\to b} = \dfrac{N_{r\to b}(\lambda')}{N_{r\to b}(\lambda') + N_{r\to r}(\lambda')} .
\end{equation}
We may then compute the $\rho$'s from measured dichroic mirror transmission and reflection curves and model $N_{\text{in}}(\lambda)$. Finally, it is important to note that our red-blue color scheme is parametrized by the arbitrary cutoff wavelength $\lambda'$. We may choose $\lambda'$ to minimize the overall fraction of wrong settings,
\begin{equation}
\lambda' = \arg\min\cb{\dfrac{N_{r\to b} + N_{b\to r}} {N_{r\to r} + N_{r \to b} + N_{b \to r} + N_{b \to b}}}.
\label{eq:totalfw}
\end{equation}
For the four stars in the two observing runs, and the model of $N_{\rm in}(\lambda)$ described in the next section, the wrong-way fractions are tabulated in Table~\ref{tab:fwtable}. One typical analysis is illustrated in Fig.~\ref{fig:uberplot}.

%----------------------------------------------------------------------------------------------------------------------------------------------------------------------------------
\vspace{\skips}
\subsection{Characterizing Dichroics}
\vspace{\skips}
%----------------------------------------------------------------------------------------------------------------------------------------------------------------------------------

Our setting reader uses a system of one shortpass ($s$) (Thorlabs M 254H45) and one longpass ($l$)  (Thorlabs M 254C45) beamsplitter with transmission ($T$) and reflection ($R$) probabilities plotted in Fig.~\ref{fig:uberplot}C. We choose to place the longpass beamsplitter in the reflected arm of the shortpass beamsplitter, instead of the other way around, to minimize the overall wrong-way fraction as written in Eq.~(\ref{eq:totalfw}). With this arrangement, $\rho_{\rm blue}(\lambda) = \rho_{T,s}(\lambda) \sim 10^{-3}$ for red wavelengths and $\rho_{\rm red}(\lambda) = \rho_{R,s}(\lambda)\rho_{T,l}(\lambda) \sim 10^{-3}$ for blue wavelengths. The transmission/reflection spectra of both dichroic mirrors and of the blue/red arms are plotted in Fig.~\ref{fig:uberplot}C.

%----------------------------------------------------------------------------------------------------------------------------------------------------------------------------------
\vspace{\skips}
\subsection{Modeling the number distribution of photons}
\vspace{\skips}
%----------------------------------------------------------------------------------------------------------------------------------------------------------------------------------

In this section, we describe our model of $N_{\rm in}(\lambda)$, which covers the wavelength range 350\,nm-1150\,nm. We start with the stellar spectra, which can be modeled as blackbodies with characteristic temperatures taken from the Hipparcos catalogue~\cite{perryman97,vanleeuwen07a}. We then apply corrections for the atmospheric transmission $\rho_{\text{atm}}(\lambda)$, two layers of anti-reflection coatings in each arm $\rho_{\text{lens}}(\lambda)$, a silvered mirror $\rho_{\text{mirror}}$, and finally the detector's quantum efficiency $\rho_{\text{det}}(\lambda)$ as the photon makes its way through the setting reader.

%----------------------------------------------------------------------------------------------------------------------------------------------------------------------------------
\vspace{\skips}
\subsubsection{Stellar Spectra}
\label{sec:starspectra}
\vspace{\skips}
%----------------------------------------------------------------------------------------------------------------------------------------------------------------------------------
As discussed in the main text, the stars were selected on the basis of their brightness, with temperatures ranging from 3150\,K-7600\,K. To a very good approximation, the photons emitted by the stars follow a blackbody distribution, which we assume is largely unaltered by the interstellar medium as the light travels towards earth:
\begin{eqnarray}
N_{\rm star}(\lambda) = \dfrac{2c}{\lambda^4} \frac{1}{\left[ \exp(hc/(k_b T)) - 1 \right]} .
\label{eq:bb}
\end{eqnarray}
This blackbody distribution is used as a starting point for $N_{\rm in}(\lambda)$, to which modifications will be made. The blackbody distributions for the Run 1 stars are shown in Fig.~\ref{fig:uberplot}A.

%----------------------------------------------------------------------------------------------------------------------------------------------------------------------------------
\vspace{\skips}
\subsubsection{Atmospheric Absorbance}
\label{sec:atmo}
\vspace{\skips}
%----------------------------------------------------------------------------------------------------------------------------------------------------------------------------------
We generate an atmospheric transmittance spectrum with the MODTRAN model for mid-latitude atmospheres looking towards zenith~\cite{berk1987modtran}. To correct for the observation airmass (up to $X = 2.5$), we use optical densities from \cite{frohlich1980new} to compute the atmospheric transmission efficiency, which is due mostly to broadband Rayleigh scattering. A more sophisticated model could also compute modified absorption lines at higher airmasses, but the effect on the wrong-way fractions $f_{r\to b}, f_{b\to r}$ is negligible compared to the spectral change resulting from Rayleigh scattering.

%----------------------------------------------------------------------------------------------------------------------------------------------------------------------------------
\vspace{\skips}
\subsubsection{Lenses and Detectors}
\label{sec:lens}
\begin{figure}
\centering
\includegraphics[width=3.1in]{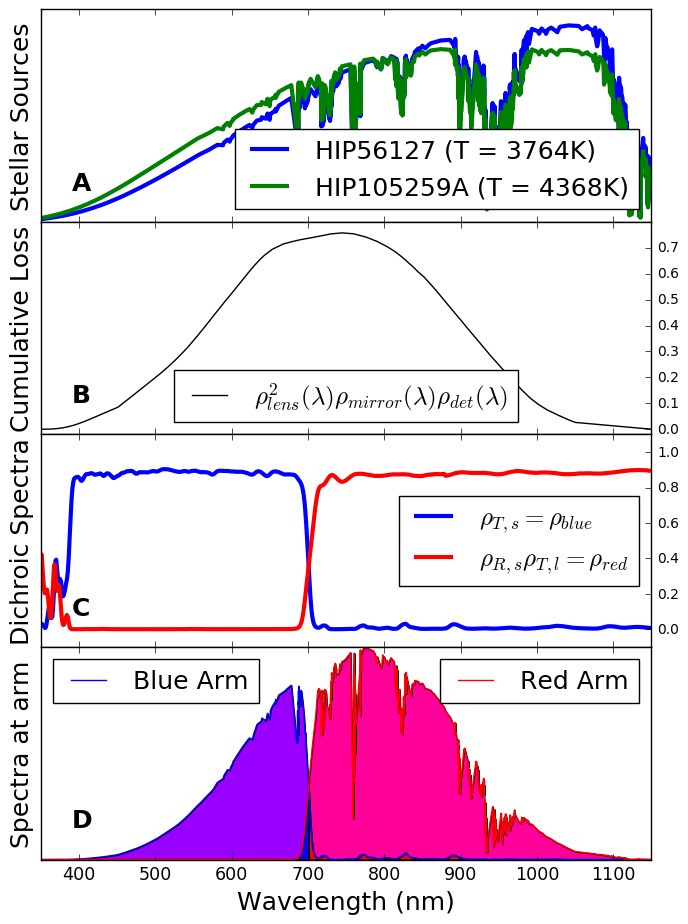}
\caption{
\small
(A) Blackbody spectra of the stars used in Run 1, extincted by atmospheric Rayleigh scattering and telluric absorption, plotted in number flux per wavelength. (B) Our maximally conservative model of the anti-reflection coatings in the two lenses, the silver mirror, and detector quantum efficiency curves as a function of photon wavelength $\lambda$. (C) Which-way probabilities as a function of $\lambda$ due to the dichroic beamsplitters. Note that the addition of the longpass beamsplitter makes $\rho_{\rm red}(\lambda)$ exceptionally flat, i.e. very good at rejecting blue photons. (D) Color distribution of photons seen at each arm are plotted, i.e. $N_{\rm in}\rho_{\rm red}$ and $N_{\rm in}\rho_{\rm blue}$. The curves are normalized so that the total area under the sum of both curves is 1. The color scheme's cutoff wavelength $\lambda^{'}$ is depicted by the shading color, and for this star is about $\lambda^{'} \sim$703.2\,nm. Note that some of the photons arriving at each arm are classified as the wrong color (overlap of red and blue arm spectra), no matter which $\lambda^{'}$ is chosen.}
\label{fig:uberplot}
\end{figure}

\begin{figure}
\centering
\includegraphics[width=3.1in]{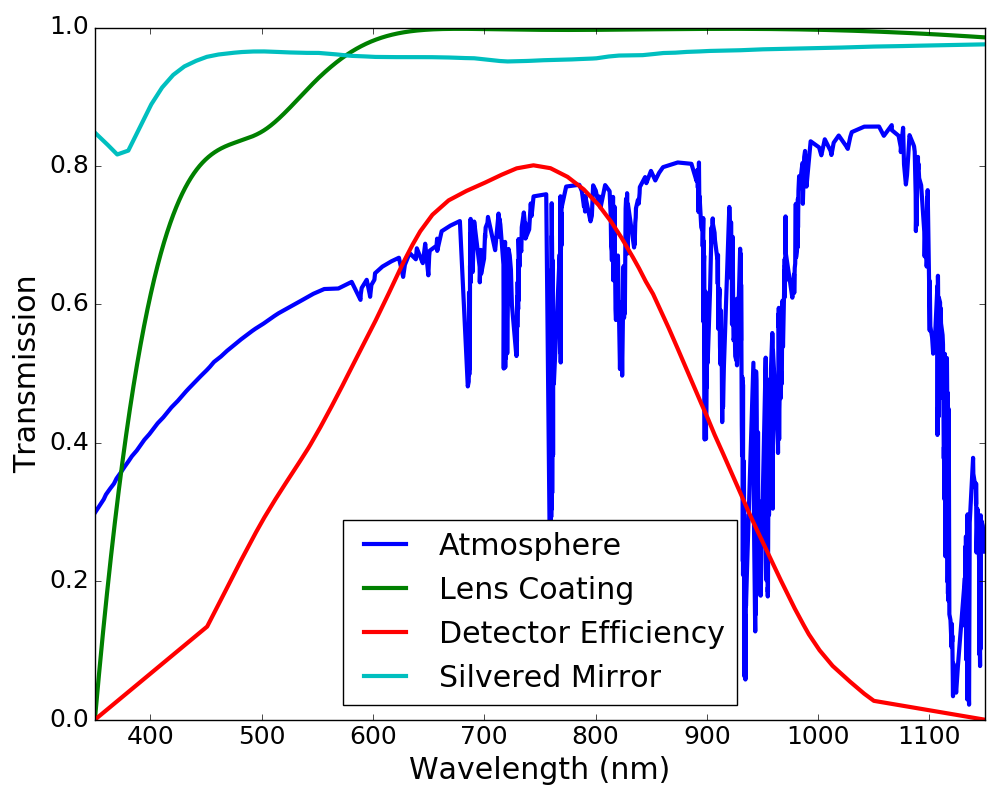}
\vspace{-0.6cm}
\caption{
\small
Wavelength-dependent probabilities of transmission through each element in the photon's path from the star to detection.}
\label{fig:transcomponents}
\end{figure}

\vspace{\skips}
%----------------------------------------------------------------------------------------------------------------------------------------------------------------------------------
In the experimental setup, one achromatic lens in each arm collimates the incident beam of stellar photons. The collimated beam reflects off a silver mirror and is focused by a second lens onto the active area of the SPADs. These elements are appropriately coated in the range from 500\,nm-1500\,nm for minimum losses. However, not all photons are transmitted through the two lenses and the mirror. Each component has a wavelength-dependent probability of transmission that is close to unity for most of the nominal range, as plotted in Fig.~\ref{fig:transcomponents}. Once the focused light is incident on the SPAD, it will actually detect the photon with some wavelength-dependent quantum efficiency. The cumulative effect of these components on the incident spectrum is shown in  Fig.~\ref{fig:uberplot}B. 

%%%%%%%%%%%%%%%%%%%%%%%%%%%%%%%%%%%%%%%%
%%%%%%%%%%%%%%%%%%%%%%%%%%%%%%%%%%%%%%%%

%%%%%%%%%%%%%%%%%%%%%%%%%%%%%%%%%%%%%%%%
%%%%%%%%%%%%%%%%%%%%%%%%%%%%%%%%%%%%%%%%

%%%%%%%%%%%%%%%%%%%%%%%%%%%%%%%%%%%%%%%%
\begin{table}
\centering
\footnotesize
\begin{tabular}{|c|c|c|c|c|c|}
\hline
\textbf{Run} & \textbf{Side} & \textbf{HIP ID} & ${f_{1\to 2}}$ & ${f_{2\to 1}}$ & \textbf{Efficiency}\\
\hline
1     & A & 56127 & 0.0142  & 0.0192 & 25.0\%\\
\hline
1     & B &  105259A & 0.0180  & 0.0146 & 24.9\%\\
\hline
2     & A & 80620 & 0.0139 & 0.0203  & 24.3\%\\
\hline
2     & B &  2876 & 0.0139 & 0.0160 & 22.7\%\\
\hline 
\end{tabular}
\caption{
\small
For each star, we compute the fraction of wrong-way photons $f$ and the atmospheric efficiency: the fraction of stellar photons directed towards our telescopes which end up generating measurement settings (as opposed to those which do not due to telluric absorption or detector inefficiencies). We adopt the notation ${f_{1\to 2}}$ and ${f_{2\to 1}}$ to allow easier indexing of the labels for the ``red" and ``blue" settings ports as applied to each run. Spectral model assumptions for other optical elements shift the $f$ values upwards by no more than $\lesssim$10\%, assuming that any uncertainties due to the atmospheric model or real-time atmospheric variability are negligible during each $\sim$ 3 minute experimental run. We therefore assume conservative values $\sigma_f/f=0.1$ in the following sections.
}
\label{tab:fwtable}
\end{table}
%%%%%%%%%%%%%%%%%%%%%%%%%%%%%%%%%%%%%%%%

%----------------------------------------------------------------------------------------------------------------------------------------------------------------------------------
\vspace{\skips}
\section{Data Analysis}
\label{sec:data}
\vspace{\skips}
%----------------------------------------------------------------------------------------------------------------------------------------------------------------------------------

In this section we analyze the data from the two experimental runs. We make the assumptions of fair sampling and fair coincidences\ \cite{larsson14}. Thus, for testing local realism, all data can be postselected to coincidence events between Alice's and Bob's measurement stations. These coincidences were identified using a time window of $2.5\,$ns.

We denote by $N_{ij}^{AB}$ the number of coincidences in which Alice had outcome $A\in\{+,-\}$ under setting $a_{i}$ ($i\!=\!1,2$) and Bob had outcome $B\in\{+,-\}$ under setting $b_{j}$ ($j\!=\!1,2$). The measured coincidences for run 1 were
\begin{equation}%
\begin{array}
[c]{lrrrr}%
ij\;\backslash\;AB & ++ & +- & -+ & --\vspace{0.1cm}\\
11 & 2\,495 & 6\,406 & \;7\,840 & 2\,234 \\
12 & 6\,545 & \;24\,073 & 30\,223 & 4\,615\\
21 & 1\,184 & 4\,537 & 5\,113 & 959\\
22 & \;18\,451 & 3\,512 & 3\,949 & \;14\,196
\end{array}
\label{eq coinc}%
\end{equation}
The coincidence numbers for run 2 were
\begin{equation}%
\begin{array}
[c]{lrrrr}%
ij\;\backslash\;AB & ++ & +- & -+ & --\vspace{0.1cm}\\
11 & 3\,703 & 10\,980 & \;14\,087 & 2\,756\\
12 & 3\,253 & \;12\,213 & 15\,210 & 2\,899\\
21 & 1\,084 & 4\,105 & 5\,442 & 932\\
22 & \;5\,359 & 1\,012 & 1\,249 & \;4\,495
\end{array}
\label{eq coinc 2}
\end{equation}

We can define the number of all coincidences for setting combination
$a_{i}b_{j}$,%
\begin{equation}
N_{ij}\equiv%
{\displaystyle\sum\nolimits_{A,B=+,-}}
N_{ij}^{AB},
\end{equation}
and the total number of all recorded coincidences,
\begin{equation}
N\equiv%
{\displaystyle\sum\nolimits_{i,j=1,2}}
N_{ij}.
\label{Ntrials}
\end{equation}
A point estimate gives the joint setting choice probabilities
\begin{equation}
q_{ij}\equiv p(a_{i}b_{j})=\frac{N_{ij}}{N}. 
\label{eq qij}
\end{equation}
We first test whether the probabilities $q_{ij}$ can be factorized, i.e., that they can be (approximately) written as%
\begin{equation}
p_{ij}\equiv p(a_{i})\,p(b_{j}).
\end{equation}
Otherwise, there could be a common cause and the setting choices would not be independent. We define $p(a_i) \equiv (N_{i1}+N_{i2})/N$ and $p(b_j) \equiv (N_{1j}+N_{2j})/N$. This leads to the following values for the individual setting probabilities for experimental run 1:
\begin{equation}
\begin{split}
p(a_{1})  &  =0.6193,\;\;p(a_{2})=0.3807,\label{eq p(a)}\\
p(b_{1})  &  =0.2257,\;\;p(b_{2})=0.7743. 
\end{split}
\end{equation}
Pearson's $\chi^{2}$-test for independence, $q_{ij}=p_{ij}$, yields $\chi^{2}=N\,%
{\textstyle\sum\nolimits_{i,j=1,2}}
\,(q_{ij}\!-\!p_{ij})^{2}/p_{ij}=1.132$. This implies that, under the assumption of independent setting choices (\ref{eq p(a)}), there is a purely statistical chance of $0.287$ that the observed data $q_{ij}$ (or data even more deviating) are obtained. This probability is much larger than any typically used threshold for statistical significance. Hence, this test does not allow a refutation of the assumption of independent setting choices. For run 2, we estimate $p(a_{1})=0.7333$, $p(a_{2})=0.2667$, $p(b_{1})=0.4854$, and $p(b_{2})=0.5146$, with $\chi^{2}=1.158$ and statistical chance $0.282$.

We next estimate the conditional probabilities for correlated outcomes in which both parties observe the same result:
\begin{equation}
p(A\!=\!B|a_{i}b_{j})=\frac{N_{ij}^{++}+N_{ij}^{--}}{N_{ij}}.
\label{eq cond}
\end{equation}
The Clauser-Horne-Shimony-Holt (CHSH) inequality\ \cite{clauser69} can be
written as
\begin{align}
C\equiv &  -p(A\!=\!B|a_{1}b_{1})-p(A\!=\!B|a_{1}b_{2})\nonumber\\
&  -p(A\!=\!B|a_{2}b_{1})+p(A\!=\!B|a_{2}b_{2})\leq0. 
\label{eq C}%
\end{align}
While the local-realist bound is $0$, the quantum bound is $\sqrt{2}\!-\!1=0.414$, and the logical (algebraic) bound is $1$.

\begin{figure}[t]
\begin{center}
\includegraphics[width=0.45\textwidth]{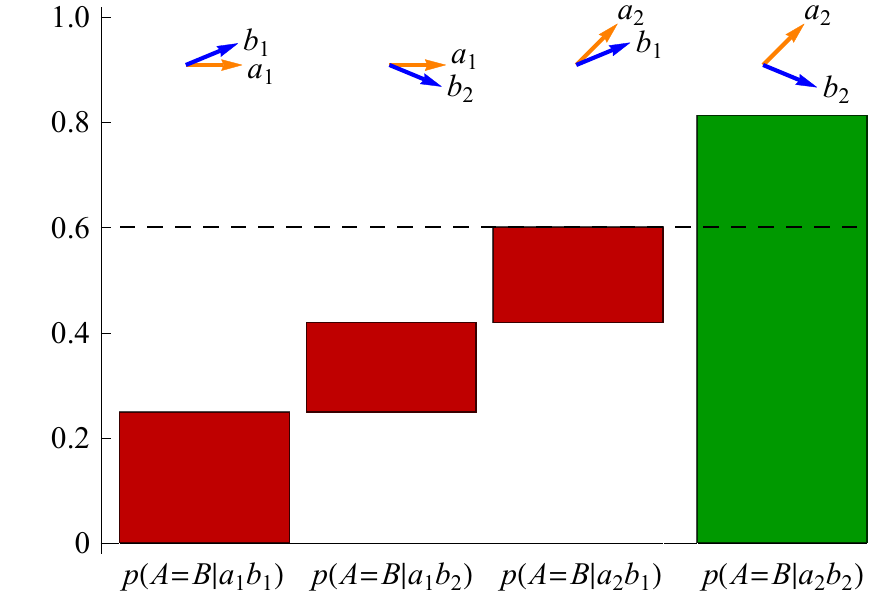}
\end{center}
\vspace{-0.6cm}
\caption{\small Plot of the conditional probabilities in Eq.~(\ref{eq C}), the CHSH inequality, calculated for run 1. The three negative contributions (red) are outweighed by the positive contribution (green), violating the local-realist bound. The red contributions are of unequal size due to limited state visibility and imperfect alignment of the polarization settings.
}
\label{Fig_Historgram}
\end{figure}

With our data, the CHSH values are $C=0.2125$ for run 1, and $C=0.2509$ for run 2, in each case violating the local-realist bound of zero. See Fig.\ \ref{Fig_Historgram}. The widely known CHSH expression in terms of correlation functions, $S\equiv|E_{11}+E_{12}+E_{21}-E_{22}|\leq2$ with $E_{ij}=2\,p(A\!=\!B|a_{i}b_{j})-1$, yields $S=2\,|\!-\!C-1|=2.425$ for run 1 and $S=2.502$ for run 2, violating the corresponding local-realist bound of $2$.

%%%%%%%%%%%%%%%%%%%%%%%%%%%%%%%%%%%%%%%%
%%%%%%%%%%%%%%%%%%%%%%%%%%%%%%%%%%%%%%%%

%----------------------------------------------------------------------------------------------------------------------------------------------------------------------------------
\vspace{\skips}
\subsection{Predictability of Settings}
\vspace{\skips}
%----------------------------------------------------------------------------------------------------------------------------------------------------------------------------------

We need to take into account two sources of imperfections in the experiment that can lead to an excess predictability\ \cite{kofler16} of the setting choices. The excess predictability $\epsilon$ quantifies the fraction of runs in which --- given all possible knowledge about the setting generation process that can be available at the emission event of the particle pairs and thus at the distant measurement events --- one could predict a specific setting better than what would simply be inferred from the overall bias of the setting choices. Loosely speaking, $\epsilon$ quantifies the fraction of runs in which the locality and freedom-of-choice assumptions fail.

%%%%%%%%%%%%%%%%%%%%%%%%%%%%%%%%%%%%%%%%
\linespread{1.0}
\begin{table}[ht]
\footnotesize
\centering
\begin{tabular}{| c | c | r | r | r | r | c | } 
\hline
Run & Side & 
\multicolumn{1}{ c |}{$r_{a_{1}}, r_{b_{1}}$} & 
\multicolumn{1}{ c |}{$r_{a_{2}}, r_{b_{2}}$} & 
\multicolumn{1}{ c |}{$n_{a_{1}}, n_{b_{1}}$} & 
\multicolumn{1}{ c |}{$n_{a_{2}}, n_{b_{2}}$} & 
\multicolumn{1}{ c |}{$\Delta t_{n_{k}}\,$[s]}  \\
\hline
$1$ &  $A$ & $105571 \pm 25$  & $38743 \pm 15$ & $1802 \pm 6$   & $1313 \pm 5$ & $59$ \\
       &  $B$ &  $26849 \pm 13$  & $93045 \pm 23$ & $756 \pm 4$ &   $1008 \pm 5$ & $59$ \\
%$1$ &  $A$ & $107\,827 \pm 25$  & $70\,161 \pm 20$ & $1\,841 \pm 6$ & $2\,377 \pm 7$ & $59$ \\
%    &  $B$ &  $31\,607 \pm 14$  & $97\,608 \pm 24$ & $890 \pm 4$    & $1\,057 \pm 5$ & $59$ \\
\hline
$2$ & $A$  & $104999 \pm 25$  & $38176 \pm 15$ & $1658 \pm 8$ & $1823 \pm 8$ & $29$ \\
    & $B$  & $59513 \pm 19$   & $67880 \pm 20$ & $1731 \pm 8$ & $1414 \pm 7$ & $30$ \\
%$2$ & $A$  & $114\,910 \pm 26$  & $48\,619 \pm 17$ & $1\,815 \pm 8$ & $2\,322 \pm 9$ & $29$ \\
%    & $B$  & $67\,383 \pm 20$   & $75\,660 \pm 21$ & $1\,959 \pm 9$ & $1\,576 \pm 8$ & $30$ \\
\hline
\end{tabular}\par
\vspace{-0.2cm}
\caption{
\small
For runs 1 and 2, $r_{k_{i}}$ and $n_{k_{i}}$ are the total and noise rates in Hz for observer $k=\{a,b\}$ and setting port $i=\{1,2\}$. We use Poisson process standard deviations $\sigma_{r_{k_{i}}} \approx \sqrt{ r_{k_{i}} / \Delta t_{r_{k}}}$, and $\sigma_{n_{k_{i}}} \approx \sqrt{ n_{k_{i}} / \Delta t_{n_{k}}}$, to estimate total and noise rate uncertainties (rounded up to the nearest integer). $\Delta t_{r_{k}}=179$\,s is the duration of both runs 1 and 2 used to measure the total rate $r_{k_{i}}$ for both observers. $\Delta t_{n_{k}}$ are the durations of control measurements to obtain the noise rates $n_{k_{i}}$ for Alice and Bob in each run. Different surface temperatures and apparent magnitudes of the stars result in different emitted spectra and thus in different count rates for run 1 and 2.}
\label{tab:exp2}
\end{table}
%%%%%%%%%%%%%%%%%%%%%%%%%%%%%%%%%%%%%%%%

The first source of imperfection is that not all of Alice's and Bob's settings were generated by photons from the two distant stars but were due to other, much closer ``noise" sources. The total rates of photons in the respective setting generation ports for runs 1 and 2 are listed in Table~\ref{tab:exp2}. Note that if one calculated $p(a_{i})$ as $r_{a_{i}}/(r_{a_{1}}\!+\!r_{a_{2}})$ and analogously for $b_{j}$, the numbers would be slightly different than the numbers in Eq.~(\ref{eq p(a)}) inferred from the coincidences. The reason is that the average duration for which a setting was valid depended slightly on the setting itself. The overall setting validity times for the whole runtime of the experiment match the numbers in Eq.~(\ref{eq p(a)}) very well.

A control measurement, pointing the receiving telescopes marginally away from the stars, yielded the noise rates listed in Table~\ref{tab:exp2}. In the most conservative case, one would assume that all noise photons were under the control of a local hidden-variable model. Thus, their contribution to the predictability of setting $a_{1}$ ($a_{2}$) would be given by the ratio of noise rate to total rate, $n_{a_{1}}/r_{a_{1}}\!=\!0.017$ ($n_{a_{2}}/r_{a_{2}}\!=\!0.034$) for run 1. Similarly, the noise contribution to the predictability for $b_{1}$ ($b_{2}$) is given by $n_{b_{1}}/r_{b_{1}}\!=\!0.028$ ($n_{b_{2}}/r_{b_{2}}\!=\!0.011$) for run 1.

The second source of imperfection is that a certain fraction of stellar photons leaves the dichroic mirror in the wrong output port. We index the wrong-way fractions $f_{i^{\prime} \rightarrow i}$ as defined in Table~\ref{tab:fwtable} with $i^{\prime} \rightarrow i$ denoting either $1 \rightarrow 2$ or $2 \rightarrow 1$.

With $(A)$ and $(B)$ denoting Alice and Bob, we can write
\begin{equation}
\begin{split}
r_{a_{i}}  &  =\left(  1-f_{i\rightarrow i^{\prime}}^{(A)}\right)
s_{i}^{(A)}+f_{i^{\prime}\rightarrow i}^{(A)}\,s_{i^{\prime}%
}^{(A)}+n_{a_{i}},
\\
r_{b_{j}}  &  =\left(  1-f_{j\rightarrow j^{\prime}}^{(B)}\right)
s_{j}^{(B)}+f_{j^{\prime}\rightarrow j}^{(B)}\,s_{j^{\prime}%
}^{(B)}+n_{b_{j}}. 
\label{eq rai}
\end{split}
\end{equation}
Here $s_{i}^{(A)}$ ($s_{j}^{(B)}$) is the detected rate of stellar photons at Alice (Bob) which have a color that, when correctly identified, leads to the setting choice $a_{i}$ ($b_{j}$). Each rate in Eq.~(\ref{eq rai}) is a sum of three terms:\ correctly identified stellar photons, incorrectly identified stellar photons that should have led to the other setting, and the noise rate. The four expressions in Eq.~(\ref{eq rai}) allow us to
find the four rates $s_{i}^{(A)}$ and $s_{j}^{(B)}$ as functions of the $f$ parameters.

We now want to quantify the setting predictability due to the dichroic mirror errors. We imagine a hidden-variable model with arbitrary local power with the following restrictions: It cannot use non-detections to its advantage, and it can only alter at most certain fractions of the incoming stellar photons, which are quantified by the dichroic mirror error probabilities. We first focus only on Alice's side. We assume that in a certain fraction of runs the local-realist model `attacks' by enforcing a specific setting value and choosing hidden variables that optimize the measurement results to maximize the Bell violation. This could for instance happen with a hidden (slower than light) signal from the source to Alice's dichroic mirror. Let us assume that $q_{a_{i}}$ is the fraction of runs in which the model decides to generate setting $a_{i}$. If the incoming stellar photon would, under correct identification, have led to setting $a_{i^{\prime}}$, this `overruling' gets reflected in the dichroic mirror error probability $f_{i^{\prime}\rightarrow i}^{(A)}$. In fact, we can equate $q_{a_{i}}=f_{i^{\prime}\rightarrow i}^{(A)}$, as the commitment to enforce setting $a_{i}$ to occur, independent of knowledge of the incoming photon's wavelength.  Thus, the probability to enforce $a_{i}$, $q_{a_{i}}$, is identical to the conditional probability $f_{i^{\prime}\rightarrow i}^{(A)}$ that $a_{i}$ is enforced although $a_{i^{\prime}}$ would have been generated otherwise. The predictability from this `overruling' is quantified by $f_{i^{\prime}\rightarrow i}^{(A)}\,s_{i^{\prime}}^{(A)}/r_{a_{i}}$, i.e.\ the fraction of $a_{i}$ settings which stem from stellar photons that should have led to setting $a_{i^{\prime}}$.

On the other hand, if the incoming stellar photon would have led to setting $a_{i}$ anyway, there is no visible `overruling' and the attack remains hidden, while the model still produces outcomes that maximize the Bell violation. The predictability from this is quantified by $f_{i^{\prime}\rightarrow i}^{(A)}\,s_{i}^{(A)}/r_{a_{i}}$, i.e.\ the fraction of $a_{i}$ settings for which no attack was actually necessary to maximize Bell violation.

Everything is analogous for Bob. In total, we can add up the different contributions---noise photons and dichroic mirror wrong-way fractions---and obtain the excess predictabilities
\begin{equation}
\begin{split}
\epsilon_{a_{i}}  &  =\tfrac{1}{r_{a_{i}}}\left(  n_{a_{i}}+f_{i^{\prime}\rightarrow i}^{(A)}\,s^{(A)}\right)  ,\\
\epsilon_{b_{j}}  &  =\tfrac{1}{r_{b_{j}}}\left(  n_{b_{j}}+f_{j^{\prime}\rightarrow j}^{(B)}\,s^{(B)}\right)  ,
\end{split}
\label{epsilonab}
\end{equation}
with the total detected stellar photon rates $s^{(A)}\equiv s_{1}^{(A)}+s_{2}^{(A)}$ and $s^{(B)}\equiv s_{1}^{(B)}+s_{2}^{(B)}$. Note that $s^{(A)} = \sum_{i} (r_{a_{i}} - n_{a_{i}})$ and $s^{(B)} = \sum_{j} (r_{b_{j}} - n_{b_{j}})$, such that the total star counts from both ports for Alice or Bob are themselves independent of any $f$ parameters, since {\it all} detected stellar photons must either go to the correct or incorrect port.

One final source that can contribute to the excess predictability concerns the physical response of the setting readers: after one of the detectors clicks with a certain setting (for example, upon detecting a red stellar photon), that detector becomes ``blind" for a dead time of approximately $500$ ns, after which its quantum efficiency recovers to the original value. During this dead / recovery time, the `blue' detector is more likely to click. Such situations would yield an excess predictability, over and above the likelihood that a hidden-variable mechanism might discern from the biased settings frequencies (unequal $q_{ij}$) or the other sources of noise and errors in the settings readers (nonzero $\epsilon_{a_i}, \epsilon_{b_j}$).

To address this additional predictability from the dead/recovery time of the setting readers, we introduced an additional, artificial ``dead time" for the `blue' detector, after the corresponding `red' detector had clicked (and vice versa). We optimized the window $\tau_{\rm cut}$ for each detector by analyzing data from our calibration measurements with the various astronomical sources, conducted before each experimental run. Then we post-selected (and deleted) all measurement coincidences from our Bell-test data that had a `blue' click within $\tau_{\rm cut}$ of a `red' click (and vice versa), consistent with the assumption of ``fair sampling" and ``fair coincidences." By finding optimal values of $\tau_{\rm cut}$ for each detector and each experimental run, we may reduce this additional, ``dead-time" predictability to an arbitrarily small amount. The effect is to remove any additional correlations between neighboring setting-detector `clicks,' beyond what would be inferred from the measured bias and $\epsilon$ predictability.

In the worst case, the predictable setting events do not happen simultaneously on both sides but fully add up. Hence, the fraction of predictable coincidences within the ensemble of setting combination $a_{i}b_{j}$ is at most
\begin{equation}
\epsilon_{ij}\equiv\epsilon_{a_{i}}+\epsilon_{b_{j}}.
\label{epsilonij}
\end{equation}
If this number is larger than $1$, $\epsilon_{ij}$ is set to $1$. 

%%%%%%%%%%%%%%%%%%%%%%%%%%%%%%%%%%%%%%%%
\linespread{1.0}
\begin{table}
\centering
\footnotesize
\begin{tabular}{| c | c | c | c | c |}
\hline
\textbf{Run} & \multicolumn{1}{ c |}{ $\epsilon_{11} \pm \sigma_{\epsilon_{11}}$  }  & \multicolumn{1}{ c |}{ $\epsilon_{12} \pm \sigma_{\epsilon_{12}}$  }  & \multicolumn{1}{ c |}{ $\epsilon_{21} \pm \sigma_{\epsilon_{21}}$  }  & \multicolumn{1}{ c |}{ $\epsilon_{22} \pm \sigma_{\epsilon_{22}}$  } \\
\hline
\multirow{ 2}{*}{ 1}      & 0.13521                              & 0.07645                             & 0.17791                              & 0.11915    \\
                                   & $\pm 6.92 \times 10^{-3}$  & $\pm 3.44 \times 10^{-3}$  & $\pm 8.24 \times 10^{-3}$  & $\pm 5.65 \times 10^{-3}$  \\
%\multirow{ 2}{*}{ 1}      & 0.13503                             & 0.08233                             & 0.15592                             & 0.10323 \\
%                                   & $\pm 6.65 \times 10^{-3}$ & $\pm 3.89 \times 10^{-3}$ & $\pm 6.85 \times 10^{-3}$ & $\pm 4.22 \times 10^{-3}$ \\
\hline
\multirow{ 2}{*}{ 2}      & 0.10533                             & 0.08917                              & 0.16094                              & 0.14477   \\
                                   & $\pm 4.30 \times 10^{-3}$  & $\pm 3.72 \times 10^{-3}$  & $\pm 6.08 \times 10^{-3}$  & $\pm 5.68 \times 10^{-3}$  \\
%\multirow{ 2}{*}{ 2}      & 0.10620                              & 0.09051                            & 0.15538                             & 0.13968 \\
%                                   & $\pm 4.35 \times 10^{-3}$ & $\pm 3.82 \times 10^{-3}$ & $\pm 5.62 \times 10^{-3}$ & $\pm 5.22 \times 10^{-3}$ \\
\hline 
\end{tabular}
\caption{
\small
For runs 1 and 2, we compute $\epsilon_{ij}$ with Eq.~(\ref{epsilonij}) and $\sigma_{\epsilon_{ij}}$ with Eqs.~(\ref{sigmaepsilonab})-(\ref{sigmaepsilonij}). We use values and errors on the total and noise rates from Table~\ref{tab:exp2} along with 10\% fractional uncertainties on the dichroic mirror wrong-way fractions in Table~\ref{tab:fwtable}. For both runs, Eqs.~(\ref{epsilon}) and (\ref{sigmaepsilon}) yield $\epsilon \pm \sigma_{\epsilon} = \epsilon_{21} \pm \sigma_{\epsilon_{21}}$.
}
\label{tab:epsijtable}
\end{table}
%%%%%%%%%%%%%%%%%%%%%%%%%%%%%%%%%%%%%%%%

We conservatively assume that all predictable events are maximally exploited by a local hidden-variable model. Then, in fact, the largest of the four fractions, i.e.,%
\begin{equation}
\epsilon\equiv\max\nolimits_{ij}\epsilon_{ij} \equiv \max\nolimits_{i} \epsilon_{a_{i}} + \max\nolimits_{j} \epsilon_{b_{j}}  ,
\label{epsilon}
\end{equation}
can be reached for the CHSH expression $C$.

To make this clear, let us consider the simple hidden-variable model in which the outcome values are always $A_{1}=-1$, $A_{2}=+1$, $B_{1}=+1$, $B_{2}=+1$, with subscripts indicating the respective setting. The first two probabilities in Eq.~(\ref{eq C}) are each $0$ (only anticorrelations), the last two are each $1$ (only correlations), and $C=0$. Now, if in a fraction $\epsilon_{21}$ of all coincidence events with setting combination $a_{2}b_{1}$ there is setting information of one party available at the source or the distant measurement event, then that latter measurement outcome can be `reprogrammed' to produce an anticorrelation. Hence, we have $p(A\!=\!B|a_{2}b_{1})=0$ in that $\epsilon_{21}$ subensemble, and $p(A\!=\!B|a_{2}b_{1})=1-\epsilon_{21}$ in total. This leads to $C=\epsilon_{21}$. Similar examples can be constructed for the other fractions. The predictabilities $\epsilon_{ij}$ thus require us to adapt the CHSH inequality of Eq.~(\ref{eq C}) to (see Ref.~\cite{kofler16})%
\begin{equation}
C\leq\epsilon. \label{eq Ceps}%
\end{equation}

The dichroic mirror errors were characterized, taking into account the spectra of the stars and all optical elements. Using the values for $f^{(A,B)}_{i \rightarrow i'}$ in Table~\ref{tab:fwtable} and the total and noise rates from Table~\ref{tab:exp2} yields a predictability of $\epsilon=0.1779$ for run 1, such that our observed value $C=0.2125$ still represents a violation of the adapted inequality of Eq.~(\ref{eq Ceps}). Likewise for run 2, we find $\epsilon=0.1609$, again yielding $C=0.2509 > \epsilon$. See Table~\ref{tab:epsijtable}.

%----------------------------------------------------------------------------------------------------------------------------------------------------------------------------------
\vspace{\skips}
\subsubsection{Uncertainty on the Settings Predictability}
\vspace{\skips}
%----------------------------------------------------------------------------------------------------------------------------------------------------------------------------------
 
We temporarily drop the labels for Alice and Bob. Assuming that the rates $r_{i}$, $n_{i}$, and the $f$ parameters are independent (which follows from our assumption of fair sampling for all detected photons), error propagation of Eq.~(\ref{epsilonab}) yields an uncertainty estimate for $\epsilon_{i}$ given by
\begin{eqnarray}
\sigma^2_{\epsilon_{i}} & = & r^{-4}_{i} \Bigg\{ r^2_i \Big[ s^2  \sigma^2_{f_{i^{\prime}\rightarrow i}} +  \Big(1-f_{i^{\prime}\rightarrow i}\Big)^2 \sigma^2_{n_{i}} + f^2_{i^{\prime}\rightarrow i}\Big(\sigma^2_{r_{i^{\prime}}} + \sigma^2_{n_{i^{\prime}}}\Big) \Big] \nonumber \\
& + & \Big[ n_i(1 - f_{i^{\prime}\rightarrow i}) + f_{i^{\prime}\rightarrow i}\Big(r_{i^{\prime}} - n_{i^{\prime}} \Big) \Big]^2 \sigma^2_{r_{i}} \Bigg\} , 
\label{sigmaepsilonab}
\end{eqnarray}
where we note that $s = r_1 + r_2 - n_1 - n_2$. Eq.~(\ref{sigmaepsilonab}) holds for Alice or Bob by applying appropriate labels. 
If we assume Alice and Bob's predictability contributions from Eq.~(\ref{epsilonij}) are independent, we find
\begin{equation}
\sigma_{\epsilon_{ij}} = \sqrt{ \sigma^2_{\epsilon_{a_{i}}} + \sigma^2_{\epsilon_{b_{j}}} } ,
\label{sigmaepsilonij}
\end{equation}
with an estimated uncertainty on $\epsilon$ from Eq.~(\ref{epsilon}) of
\begin{equation}
\sigma_{\epsilon} = \sqrt{ \sigma^2_{\max_i \epsilon_{a_{i}}} + \sigma^2_{\max_j \epsilon_{b_{j}}} },
\label{sigmaepsilon}
\end{equation}
where $\sigma_{\max_i \epsilon_{a_{i}}}$ is the uncertainty from Eq.~(\ref{sigmaepsilonab}) on the term which maximizes $\epsilon_{a_{i}}$, and likewise for Bob. For both runs 1 and 2, assuming values and errors on the total and noise rates from Table~\ref{tab:exp2}, wrong-way fractions $f$ from Table~\ref{tab:fwtable} with conservative fractional errors of $\sigma_f/f=0.1$, Table~\ref{tab:epsijtable} shows values of $\epsilon_{ij}$ from Eqs.~(\ref{epsilonab})-(\ref{epsilonij}), $\sigma_{\epsilon_{ij}}$ from Eqs.~(\ref{sigmaepsilonab})-(\ref{sigmaepsilonij}), and $\sigma_{\epsilon}$ from Eq.~(\ref{sigmaepsilon}).

%----------------------------------------------------------------------------------------------------------------------------------------------------------------------------------
\vspace{\skips}
\subsection{Statistical significance}
\vspace{\skips}
%----------------------------------------------------------------------------------------------------------------------------------------------------------------------------------

There exist several different ways to estimate the statistical significance for experimental runs 1 and 2. The result of any such statistical analysis is a $p$-value, i.e., a bound for the probability that the null hypothesis --- local realism with $\epsilon$ predictability, biased detector-setting frequencies, fair sampling, fair coincidences, {\it and any other additional assumptions} --- could have produced the experimentally observed data by a random variation.

Until recently, it was typical in the literature on such Bell tests to estimate a $p$-value under several assumptions (e.g., \cite{scheidl10}): that each trial was independent and identically distributed (i.i.d.), and that the hidden-variable mechanism could not make any use of ``memory" of the settings and outcomes of previous trials. Under those assumptions, one typically applied Poisson statistics for single coincidence counts, and assumed that the underlying statistical distribution was Gaussian. Moreover, it was typical to neglect the excess predictability, $\epsilon$. Applied to our experimental data, such methods yield what we consider to be overly optimistic estimates, suggesting violation of the CHSH inequality by $\nu \geq 39.8$ and $42.7$ standard deviations for runs 1 and 2, respectively.

However, such an approach assumes that the measured coincidence counts $N_{ij}^{AB}$ are equal to their expected values, but then contradicts this assumption by calculating the probability that the $N_{ij}^{AB}$ could have values differing by several standard deviations from their expected values. Plus, as recent work has emphasized (e.g., \cite{kofler16}), excess predictability $\epsilon$ must be taken into account when estimating statistical significance for any violations of the CHSH inequality.

More recently, several authors have produced improved methods for calculating $p$-values for Bell tests. These newer approaches do not assume i.i.d.~trials, and also, more conservatively, allow the hidden-variable model to exploit ``memory" of previous settings and outcomes. Whereas the ``memory loophole" cannot achieve Bell violation, incorporating possible memory effects does require modified calculations of statistical significance \cite{gill03,gill14,bierhorst15,kofler16,elkous16}. 

Although these new works represent a clear advance in the literature, unfortunately they are not optimized for use with our particular experiment. For example, the unequal settings probabilities (bias) for our experiment limit the utility of the bounds derived in \cite{bierhorst15,elkous16}, as the resulting $p$-values are close to 1. Likewise, one may follow the approach of \cite{gill03,gill14,kofler16} and use the Hoeffding inequality \cite{hoeffding63}. However, it is known in general that such bounds routinely overestimate $p$ --- and hence underestimate the genuine statistical significance of a given experiment --- by a substantial amount (see, e.g., \cite{elkous16}).

Therefore, in this section we present an {\it ab initio} calculation of the $p$-value tailored more specifically to our experiment. This method yields what we consider to be reasonable upper bounds on the $p$-values, which are still highly significant even with what we regard as a conservative set of assumptions. Our calculation incorporates predictability of settings and allows the local-realist hidden-variable theory to exploit memory of previous detector settings and measurement outcomes. We present essential steps in the calculation here, and defer fuller discussion to future work.

We consider a quantity $W$, which is a weighted measure of the number of ``wins," that is, the number of measurement outcomes that contribute positively to the CHSH quantity $C$, defined in Eq.~(\ref{eq C}). A win consists of $A = B$ for settings pair $a_2b_2$, and $A \neq B$ for any other combination of settings. Thus we define $N_{ij}^{\rm win} \equiv [N_{11}^{A \neq B}, N_{12}^{A \neq B}, N_{21}^{A \neq B}, N_{22}^{A = B}]$, and
%%%%%%%%%%
\begin{equation}
W = \sum_{ij} \frac{ N_{ij}^{\rm win} }{q_{ij} ( 1 - \epsilon_{ij} ) },
\label{Wdef}
\end{equation}
where $q_{ij} \equiv N_{ij} / N$ is the fraction of trials in which settings combination $ij$ occurs, and $\epsilon_{ij}$, defined in Eq.~(\ref{epsilonij}), is the probability that a given trial will be ``corrupt." A trial is considered ``corrupt" if it (1) involved a noise (rather than stellar) photon, or (2) involved a dichroic mirror error, or (3) was previewed by the hidden-variable theory for the purpose of considering a dichroic mirror error, but was passed over because the stellar photon already had the desired color. The occurrence of a corruption in any trial is taken to be an independent random event, which has probability $\epsilon_{ij}$ that depends on the settings pair $a_i b_j$. We assume that for ``uncorrupt" trials, the hidden-variable theory has no information about what the settings pair will be beyond the probabilities $q_{ij}$.

We assume that the hidden-variable theory can exploit each corrupted trial and turn it into a win. We further assume that the occurrence of these corrupt events cannot be influenced by either the experimenter or the hidden-variable theory; they occur with uniform probability $\epsilon_{ij}$ in each trial. We consider the probabilities $\epsilon_{ij}$ to be known (to within some uncertainty $\sigma_{\epsilon_{ij}}$), but the actual number of corrupt trials to be subject to statistical fluctuations. 

The $p$-value is the probability that a local-realist hidden-variable theory, using its best possible strategy, could obtain a value of $W$ as large as the observed value. To define this precisely, we must be clear about the ensemble that we are using to define probabilities. It is common to attempt to describe the ensemble of all experiments with the same physical setup and the same number of trials. Yet it is difficult to do this in a precise way, because one has to use the statistics of settings choices observed in the experiment to determine the probabilities for the various settings. From a Bayesian point of view, this requires the assumption of a prior probability distribution on settings probabilities, and the answers one finds for $p$ would depend on what priors one assumes. 

We avoid such issues by considering the actual number $N_{ij}$ of the occurrences of each settings choice $a_i b_j$ as given. The relevant ensemble is then the ensemble of all possible {\it orders} in which the settings choices could have occurred. The $p$-value will then be the fraction of orderings for which the hidden-variable theory, using its best strategy, could obtain a value of $W$ greater than or equal to the value obtained in the experiment.  

We may motivate the form of $W$ in Eq.~(\ref{Wdef}) as follows. In the absence of noise or errors, the hidden-variable model could specify which outcomes ($A, B$) will arise for each of the possible settings ($i, j$). The best plans will win for three of the four possible settings pairs, but will lose for one of the possible settings pairs. Hence a plan may be fully specified by identifying which settings pair will be the loser. (There will actually be two detailed plans for such a specification, related by a reversal of all outcomes, but we may treat such plans as equivalent.)

In the presence of noise and errors, for each time the settings pair is $a_i b_j$, there is a probability $\epsilon_{ij}$ that the trial is corrupt. If the trial is corrupt, it automatically registers as a win. If it is not corrupt, then it has a probability $P_{ij}^{\rm win}$ of registering as a win, where we take $P_{ij}^{\rm win}$ to be $p (A = B \vert a_i b_j)$ for $(ij) = (22)$, and $p (A \neq B \vert a_i b_j)$ for the other three cases. Then we may write
%%%%%%%%
\begin{equation}
\langle N_{ij}^{\rm win} \rangle = \left[ \epsilon_{ij} + \left( 1 - \epsilon_{ij} \right) P_{ij}^{\rm win} \right] N_{ij} ,
\label{Nijwinexp}
\end{equation}
which may be solved for $P_{ij}^{\rm win}$:
%%%%%%%%%
\begin{equation}
P_{ij}^{\rm win} = \frac{ \langle N_{ij}^{\rm win} \rangle }{N_{ij} (1 - \epsilon_{ij} ) } - \frac{ \epsilon_{ij} }{1 - \epsilon_{ij} } .
\label{Pijwin}
\end{equation}
The CHSH inequality may be written $\sum_{ij} P_{ij}^{\rm win} \leq 3$, so Eq.~(\ref{Pijwin}) implies that
%%%%%%%%%
\begin{equation}
\sum_{ij} \frac{ \langle N_{ij}^{\rm win} \rangle}{q_{ij} (1 - \epsilon_{ij} ) } \leq (3 + \bar{\epsilon} ) N ,
\label{Nijwinexp2}
\end{equation}
where we have defined 
\begin{equation}
\bar{\epsilon} = \sum_{ij} \frac{ \epsilon_{ij} } {1 - \epsilon_{ij} } \ .
\label{epsbar}
\end{equation}
The lefthand side of Eq.~(\ref{Nijwinexp2}) motivates our ansatz for $W$ in Eq.~(\ref{Wdef}).

The function $W$, which is a random variable, may be expressed in terms of a set of more elementary random variables. We label the trials by $\alpha$, so for each trial $\alpha$ there will be a set of random variables:
%%%%%%%%
\begin{equation}
\begin{split}
F_{ij}^\alpha &= \bigg\{ \begin{array}{ll} 1 & \text{ if the settings pair is} \> a_i b_j \> \text{in trial} \> \alpha  \\ 0 & \text{ otherwise} \end{array} \\
G^\alpha & = \bigg\{ \begin{array}{ll} 1 & \text{ if the trial} \>\alpha \text{ is corrupt } \\ 0 & \text{ otherwise} \end{array}  \\
U^\alpha &= \bigg\{ \begin{array}{ll} 1 & \text{ if the trial} \> \alpha \text{ is uncorrupt} \\ 0 & \text{otherwise} \end{array}
\end{split}
\label{FGU}
\end{equation}
with $G^\alpha + U^\alpha = 1$. We also define the functions
%%%%%%%%%%%%
\begin{equation}
\begin{split}
\omega_{ij}^\alpha &= \bigg\{ \begin{array}{ll} 1 & \text{if the settings pair} \> a_i b_j \> \text{in trial} \> \alpha \> \text{is a win} \\ 0 & \text{otherwise} \end{array} \\
\bar{\omega}_{ij}^\alpha &= \bigg\{ \begin{array}{ll} 1 & \text{if the settings pair} \> a_i b_j \> \text{in trial} \> \alpha \> \text{is a loss } \\ 0 & \text{otherwise} \end{array}
\end{split}
\label{omegaij}
\end{equation}
with $\omega_{ij}^\alpha + \bar{\omega}_{ij}^\alpha = 1$. Unlike the variables in Eq.~(\ref{FGU}), $\omega_{ij}^\alpha$ and $\bar{\omega}_{ij}^\alpha$ are not random; they are under the control of the hidden-variable mechanism. The square of each of the quantities in Eqs.~(\ref{FGU}) and (\ref{omegaij}) is equal to itself, since their only possible values are $0$ and $1$.

Our goal is to evaluate $\sigma_W^2 = \langle W^2 \rangle - \langle W \rangle^2$. We begin by calculating $\langle W \rangle = \sum_\alpha \langle W_\alpha \rangle$. In terms of the quantities in Eqs.~(\ref{FGU}) and (\ref{omegaij}), we may write
%%%%%%%%%
\begin{equation}
W_\alpha = \sum_{ij} \frac{ F_{ij}^\alpha \left( U^\alpha \omega_{ij}^\alpha + G^\alpha \right) }{q_{ij} (1 - \epsilon_{ij} ) }.
\label{Walpha}
\end{equation}
Since the settings are chosen randomly on each trial, we assume that all orderings of the setting choices are equally likely, and are independent of the occurrence of corruptions. This implies that $\langle F_{ij}^\alpha U^\alpha \rangle = q_{ij} (1 - \epsilon_{ij})$ and $\langle F_{ij}^\alpha G^\alpha \rangle = q_{ij} \epsilon_{ij}$, independent of $\alpha$. Then we find
%%%%%%%%%%%
\begin{equation}
\begin{split}
\langle W_\alpha \rangle &= \sum_{ij} \frac{ q_{ij} \left[ (1 - \epsilon_{ij} ) \omega_{ij}^\alpha + \epsilon_{ij} \right]}{q_{ij} (1 - \epsilon_{ij} ) } \\
&= \sum_{ij} \omega_{ij}^\alpha + \sum_{ij} \frac{ \epsilon_{ij} }{1 - \epsilon_{ij} } = 3 + \bar{\epsilon} ,
\end{split}
\label{Walphaexp}
\end{equation}
and hence 
\begin{equation}
\langle W \rangle = N (3 + \bar{\epsilon} ). 
\label{Wexp}
\end{equation}

To evaluate $\langle W^2 \rangle$ we write
%%%%%%%%%%
\begin{equation}
\langle W^2 \rangle = \sum_\alpha \sum_\beta \langle W_\alpha W_\beta \rangle = \sum_\alpha \langle W_\alpha^2 \rangle + \sum_\alpha \sum_{\beta\neq \alpha} \langle W_\alpha W_\beta \rangle .
\label{W2exp}
\end{equation}
For the first term, we have
%%%%%%%%%%
\begin{equation}
\begin{split}
\langle W_\alpha^2 \rangle &= \sum_{ij} \sum_{k\ell} \frac{ \langle F_{ij}^\alpha F_{k\ell}^\alpha (U_\alpha \omega_{ij}^\alpha + G^\alpha ) (U^\alpha \omega_{k\ell}^\alpha + G^\alpha) \rangle }{q_{ij} q_{k\ell} (1 - \epsilon_{ij} ) (1 - \epsilon_{k\ell} ) } \\
&= \sum_{ij} \frac{ \omega_{ij}^\alpha }{q_{ij} (1 - \epsilon_{ij} )} + \sum_{ij} \frac{ \epsilon_{ij} }{q_{ij} (1 - \epsilon_{ij} )^2 }  ,
\end{split}
\label{Walpha2exp}
\end{equation}
where the second line follows upon noting that $F_{ij}^\alpha F_{k\ell}^\alpha = 0$ if $(ij)\neq (k\ell)$, and using $(F_{ij}^\alpha)^2 = F_{ij}^\alpha$, $(\omega_{ij}^\alpha)^2 = \omega_{ij}^\alpha$. We therefore find
%%%%%%%%%
\begin{equation}
\sum_\alpha \langle W_\alpha^2 \rangle = N \left[ \sum_{ij} \frac{1 - f_{ij} }{q_{ij} (1  - \epsilon_{ij} ) } + \sum_{ij} \frac{\epsilon_{ij} }{q_{ij} (1 - \epsilon_{ij} )^2 } \right] ,
\label{sumWalpha2exp}
\end{equation}
where we have defined $f_{ij}$ as the fraction of trials for which the hidden-variable theory chooses $(ij)$ to be the losing settings pair. 

For the second term on the righthand side of Eq.~(\ref{W2exp}), we have
%%%%%%%%%%%%
\begin{equation}
\begin{split}
\sum_\alpha& \sum_{\beta \neq \alpha} \langle W_\alpha W_\beta \rangle \\
&= \sum_\alpha \sum_{\beta \neq \alpha} \sum_{ij} \sum_{k\ell} \frac{ \langle F_{ij}^\alpha F_{k \ell}^\beta (U^\alpha \omega_{ij}^\alpha + G^\alpha ) (U^\beta \omega_{k\ell}^\beta + G^\beta ) \rangle}{q_{ij} q_{k\ell} (1 - \epsilon_{ij} ) (1 - \epsilon_{k\ell}) } \\
&= \sum_{ij} \sum_{k\ell} \sum_\alpha \sum_{\beta \neq \alpha} \frac{ q_{ij} (N q_{k\ell} - \delta_{ij,k\ell} )}{N - 1} \\
&\quad \quad \times \frac{ \left[ (1 - \epsilon_{ij} )\omega_{ij}^\alpha + \epsilon_{ij} \right] \left[ (1 - \epsilon_{k\ell} )\omega_{k\ell}^\beta + \epsilon_{k\ell} \right]}{q_{ij} q_{k\ell} (1 - \epsilon_{ij} ) (1 - \epsilon_{k\ell} )} \\
&= T_1 + T_2,
\end{split}
\label{WalphaWbetaexp1}
\end{equation}
where $\delta_{ij,k\ell} = 1$ if $(ij) = (k\ell)$ and $0$ otherwise. (We have used the fact that for each of the $N_{ij}$ values of $\alpha$ for which $F_{ij}^\alpha = 1$, there are $N_{ij} - 1$ values of $\beta \neq \alpha$ for which $F_{ij}^\beta = 1$.) 

To further simplify Eq.~(\ref{WalphaWbetaexp1}), we first assume that the hidden-variable theory cannot exploit memory of previous settings or outcomes. In that case, we may neglect correlations between $F_{ij}^\alpha$ and $\omega_{k\ell}^\beta$, and perform a full ensemble average. (We will relax this assumption below.) Proceeding as above, the term $T_1$ may then be rewritten
%%%%%%%%
\begin{equation}
\begin{split}
T_1 &= \frac{ N}{N-1} \sum_{ij} \sum_{k\ell} \sum_\alpha \sum_{\beta \neq \alpha} \frac{ 1}{(1 - \epsilon_{ij} ) (1 - \epsilon_{k\ell} )} \\
&\quad\quad \times \left[ (1 - \epsilon_{ij} )\omega_{ij}^\alpha + \epsilon_{ij} \right] \left[ (1 - \epsilon_{k\ell} )\omega_{k\ell}^\beta + \epsilon_{k\ell} \right] \\
&=N^2 (3 + \bar{\epsilon})^2,
\end{split}
\label{T1}
\end{equation}
where we have made use of the fact that $\sum_{ij} 1 / (1 - \epsilon_{ij} ) = \sum_{ij} (1 - \epsilon_{ij} ) / (1 - \epsilon_{ij} ) + \sum_{ij} \epsilon_{ij} / (1 - \epsilon_{ij} ) = 4 + \bar{\epsilon}$. For the term $T_2$, we note that 
%%%%%%%%%%%%
\begin{equation}
\sum_\alpha \sum_{\beta \neq \alpha} \omega_{ij}^\alpha \omega_{ij}^\beta = N (1 - f_{ij} ) \left[ N (1 - f_{ij} ) - 1 \right] .
\label{omegaalphabeta}
\end{equation}
Then $T_2$ may be rewritten
%%%%%%%%%%
\begin{equation}
\begin{split}
T_2 &= - \frac{1}{N-1} \sum_{ij} \sum_{\alpha} \sum_{\beta \neq \alpha} \frac{1}{ q_{ij}(1 - \epsilon_{ij})^2} \\
&\quad\quad \times \left[ (1 - \epsilon_{ij} ) \omega_{ij}^\alpha + \epsilon_{ij} \right] \left[ (1 - \epsilon_{ij} ) \omega_{ij}^\beta + \epsilon_{ij} \right] \\
&= - \frac{N}{N-1} \sum_{ij} \bigg\{ \frac{ (1 - f_{ij} ) \left[ N (1 - f_{ij} ) - 1 \right]}{q_{ij} } \bigg\} \\
&\quad\quad - N \sum_{ij} \bigg\{ \frac{ 2 \epsilon_{ij} (1 - f_{ij} )}{q_{ij} (1 - \epsilon_{ij} ) } + \frac{\epsilon_{ij}^2}{q_{ij} (1 - \epsilon_{ij} )^2 } \bigg\} .
\end{split}
\label{T2}
\end{equation}
Following some straightforward algebra, Eqs.~(\ref{sumWalpha2exp}), (\ref{T1}), and (\ref{T2}) yield
%%%%%%%%%%
\begin{equation}
\sigma_W^2 = \frac{ N^2}{N - 1} \sum_{ij} \frac{ f_{ij} (1 - f_{ij} )}{q_{ij} } + N\sum_{ij} \frac{ f_{ij} \epsilon_{ij} }{q_{ij} (1 - \epsilon_{ij} ) } .
\label{sigmaW}
\end{equation}

The $f_{ij}$ are under the control of the hidden-variable theory, so we make the conservative assumption that the hidden-variable theory may choose the $f_{ij}$ so as to maximize $\sigma_W$. To impose the constraint that $\sum_{ij} f_{ij} = 1$, we introduce a Lagrange multiplier $\lambda$:
%%%%%%%%
\begin{equation}
\begin{split}
L &= \frac{ N^2}{ N - 1} \sum_{ij} \frac{ f_{ij} (1 - f_{ij} )}{q_{ij} } + N \sum_{ij} \frac{ f_{ij} \epsilon_{ij} }{q_{ij} (1 - \epsilon_{ij} ) } \\
&\quad\quad + \lambda \left( \sum_{ij} f_{ij} - 1 \right) .
\end{split}
\label{Ldef}
\end{equation}
Setting $\partial L / \partial f_{ij} = 0$, we find the optimum values $f_{ij}^{\rm opt} (\lambda)$. By inserting these into the normalization condition $\sum_{ij} f_{ij}^{\rm opt} = 1$, we may solve for $\lambda$, which in turn yields
%%%%%%%%%%%%
\begin{equation}
f_{ij}^{\rm opt} = \frac{1}{2} - q_{ij} + \frac{ N -1}{2N} \left[ \frac{ \epsilon_{ij} }{1 - \epsilon_{ij} }- \bar{\epsilon} q_{ij} \right] .
\label{fopt}
\end{equation}
Inserting $f_{ij}^{\rm opt}$ into Eq.~(\ref{sigmaW}) for $\sigma_W^2$, we find
%%%%%%%%%%
\begin{equation}
\begin{split}
\left( \sigma_W^{\rm opt} \right)^2 &= \frac{ N^2}{4 (N - 1 ) } \left[ \left( \sum_{ij} \frac{1}{ q_{ij} } \right ) - 4 \right] - N \bar{\epsilon} \\
&\quad\quad + \frac{ N}{4} \sum_{ij} \frac{ \epsilon_{ij}}{q_{ij} (1 - \epsilon_{ij} ) } \\
&\quad\quad - \frac{1}{4} (N - 1) \bar{\epsilon}^2 + \frac{1}{4} \sum_{ij} \frac{ ( N - \epsilon_{ij} ) \epsilon_{ij} }{q_{ij} (1 - \epsilon_{ij} )^2 } .
\end{split}
\label{sigmaWopt}
\end{equation}
For run 1, Eq.~(\ref{fopt}) yields an unphysical $f_{12} < 0$ for our data.
Upon employing a second Lagrange multiplier to ensure both that $\sum_{ij} f_{ij} = 1$ and $f_{12} \geq 0$, we find 
\begin{equation}
f^{\rm opt}_{ij}  =  \frac{1}{2} + \Bigg(\frac{N-1}{2N}\Bigg) \Bigg[ \frac{\epsilon_{ij}}{1 - \epsilon_{ij}} - \frac{q_{ij}}{1-q_{12}} \Bigg(   \frac{N}{N-1} + \bar{\epsilon} - \frac{ \epsilon_{12} }{ 1 - \epsilon_{12} } \Bigg) \Bigg]\, ,
\label{foptgt0}
\end{equation}
such that $f_{11}^{\rm opt} = 0.376$, $f_{12}^{\rm opt} = 0$, $f_{21}^{\rm opt} = 0.483$, and $f_{22}^{\rm opt} = 0.141$. For run 1, one must substitute Eq.~(\ref{foptgt0}) into Eq.~(\ref{sigmaW}) in order to find $\sigma_W^{\rm opt}$. For run 2, on the other hand, Eq.~(\ref{fopt}) yields $f_{ij} > 0 \ \forall \ (ij)$, with $f_{11}^{\rm opt} = 0.101$, $f_{12}^{\rm opt} = 0.062$, $f_{21}^{\rm opt} = 0.428$, and $f_{22}^{\rm opt} = 0.409$, allowing $\sigma_W^{\rm opt}$ to be computed with Eq.~(\ref{sigmaWopt}).

Using the values for total and noise rates ($r$, $n$) in Table~\ref{tab:exp2}, dichroic mirror wrong-way fractions ($f$) in Table~\ref{tab:fwtable}, values of $q_{ij}$ inferred from Eqs.~(\ref{Ntrials}) and (\ref{eq qij}) and the probabilities for corrupt trials $\epsilon_{ij}$ in Table~\ref{tab:epsijtable}, values for $W$, $\langle W \rangle$, and $\sigma_W^{\rm opt}$ for both runs are listed in Table~\ref{tab:Ws}. 

%%%%%%%%%%%%%%%%%%%%%%%%%%%%%%%%%%%%%%%%
\linespread{1.0}
\begin{table}[ht]
\centering
\footnotesize
\begin{tabular}{| c |  l | l | l | l | l |  }
\hline
\textbf{Run} &  \multicolumn{1}{ c |}{ $W$  }  & \multicolumn{1}{ c |}{ $\langle W \rangle$  }   &  \multicolumn{1}{ c |}{ $\sigma^{\rm opt}_W$  }   \\
\hline
1     & $5.0249 \times 10^5$ & $4.8954 \times 10^5$ & 954.3\\
%1     & $7.4522 \times 10^5$ & $7.1991 \times 10^5$ & 1137.9\\
\hline
2     & $3.3030 \times 10^5$ & $3.1754 \times 10^5$ & 682.6\\
%2     & $4.9659 \times 10^5$ & $4.7793 \times 10^5$ & 811.46\\
\hline 
\end{tabular}
\caption{
\small
For runs 1 and 2, values for $W$ and $\langle W \rangle$ from Eqs.~(\ref{Wdef}) and (\ref{Wexp}) are shown, as well as values of $\sigma^{\rm opt}_W$ from Eqs.~(\ref{sigmaW}) and (\ref{foptgt0}) for run 1 and Eq.~(\ref{sigmaWopt}) for run 2.
}
\label{tab:Ws}
\end{table}
%%%%%%%%%%%%%%%%%%%%%%%%%%%%%%%%%%%%%%%%

Next we take into account the uncertainty in the predictabilities $\epsilon_{a_i}$ and $\epsilon_{b_j}$. 
The quantity of interest is
%%%%%%%%%%
\begin{equation}
\bar{\nu} = \frac{ W - \langle W \rangle}{\sigma_W^{\rm opt} } .
\label{barnu}
\end{equation}
The quantities $W$, $\langle W \rangle$, and $\sigma^{\rm opt}_W$ all depend on $\epsilon_{a_i}$ and $\epsilon_{b_j}$, along with the $N^{AB}_{ij}$ values, which are taken as given. Therefore, we only need to propagate uncertainties on $\epsilon_{a_i}$ and $\epsilon_{b_j}$ to compute the uncertainty on $\bar{\nu}$, which we denote $\Delta_{\nu}$. 

We assume no covariance between $\epsilon_{a_i}$ and $\epsilon_{b_j}$. This again follows from our assumptions of independence for Alice and Bob as well as fair sampling for all detected photons, which implies $r_i$, $n_i$, and $f_{i'\rightarrow i}$ (the inputs to $\epsilon_{a_i}$ and $\epsilon_{b_j}$) are independent. An estimate for $\Delta_\nu$ is then given by:
\begin{equation}
\begin{split}
\Delta^2_{\nu} &= \sum_{a_i} \left( \frac{ \partial \bar{\nu} }{\partial \epsilon_{a_i} } \right)^2 \sigma_{\epsilon_{a_i} }^2 + \sum_{b_j} \left( \frac{ \partial \bar{\nu} }{\partial \epsilon_{b_j} } \right)^2 \sigma_{\epsilon_{b_j} }^2 \\
&=\sum_{a_i} \left( \frac{ \sigma_{\epsilon_{a_i} } }{\sigma_W} \right)^2 \left[ \sum_j \frac{ {\cal E}_{ij}}{q_{ij} (1 - \epsilon_{ij} )^2 }  \right]^2 \\
&\quad\quad  + \sum_{b_j} \left( \frac{ \sigma_{\epsilon_{b_j} }}{\sigma_W} \right)^2 \left[ \sum_i \frac{ {\cal E}_{ij} }{q_{ij} (1 - \epsilon_{ij} )^2 }  \right]^2 ,
\end{split}
\label{Dnupart}
\end{equation}
where 
\begin{equation}
{\cal E}_{ij} \equiv N_{ij}^{\rm win} - N q_{ij} - \left( \frac{ \bar{\nu} N}{2 \sigma_W} \right) f_{ij}^{\rm opt} ,
\label{Eijdef}
\end{equation}
and we recall from Eq.~(\ref{epsilonij}) that $\epsilon_{ij} = \epsilon_{a_i} + \epsilon_{b_j}$.
We may now compute how $\sigma_{\epsilon_{a_i}}$ and $\sigma_{\epsilon_{b_j}}$ affect the statistical significance of each run. The naive number of standard deviations $\bar{\nu}$ in Eq.~(\ref{barnu}) implicitly assumed $\sigma_{\epsilon_{a_i}} , \sigma_{\epsilon_{b_j} }=0$, and therefore $\Delta_{\nu}=0$. 
If we allow for an uncertainty in $\nu$ equal to $n$ times the 1-$\sigma$ uncertainty in $\nu$, then we should calculate the $p$-value using
\begin{equation}
\nu_n \equiv \bar{\nu} - n \Delta_{\nu} \> .
\label{nunumbar}
\end{equation}
If we choose $n$ so that $n = \nu_n$, then
\begin{equation}
\nu_n =  \frac{ \bar{\nu} }{1 + \Delta_\nu} .
\label{nupess}
\end{equation}
Assuming a Gaussian distribution for large-sample experiments, we conclude that the conditional probability that the hidden variable mechanism could achieve a value of $W$ as large as the observed value $W_{\rm obs}$, assuming that the true value of $\nu \ge \nu_n$, is given by $p_{\rm cond}=\frac{1}{2} {\rm erfc} ( \nu_n / \sqrt{2} )$. Since we chose $n=\nu_n$, if we assume Gaussian statistics for the uncertainty in $\nu$, then there is an equal probability that the true value of $\nu$ is less than $\nu_n$, in which case our analysis does not apply, and we must conservatively assume that $W$ might exceed $W_{\rm obs}$.  Thus, the $p$-value corresponding to the total probability that $W \ge W_{\rm obs}$ is bounded by $p = 2 p_{\rm cond}$.  Again assuming Gaussian statistics, we can relate $p$ to an equivalent $\nu$, by $p = \frac{1}{2} {\rm erfc} (\nu/\sqrt{2})$.  Proceeding in this way, we find the values for $\bar \nu$, $\Delta_\nu$, $\nu$, and $p$ listed in Table~\ref{tab:DWs}.

%%%%%%%%%%%%%%%%%%%%%%%%%%%%%%%%%%%%%%%%
\linespread{1.0}
\begin{table}[ht]
\centering
\footnotesize
\begin{tabular}{| c | l | l | r | l | }
   \hline
\textbf{Run} 
& \multicolumn{1}{ c |}{ $\bar \nu$ }
& \multicolumn{1}{ c |}{ $\Delta_{\nu}$  } 
& \multicolumn{1}{ c |}{ $\nu$  }  
& \multicolumn{1}{ c |}{ $p$  }  \\
\hline
1       & 13.57 &  0.79905 &  7.54 & $4.64 \times 10^{-14}$\\
%1       & 22.24 &  0.87526 &  11.86 & $1.90 \times 10^{-32}$\\
\hline
2       & 18.71 &  0.53999  & 12.15 & $5.93 \times 10^{-34}$\\
%2       & 23.00 &  0.64336  & 14.00 & $1.66 \times 10^{-44}$\\
\hline 
\end{tabular}
\caption{
\small
Values for $\bar \nu$, $\Delta_\nu$, $\nu$, and $p$ for runs 1 and 2. 
}
\label{tab:DWs}
\end{table}

%----------------------------------------------------------------------------------------------------------------------------------------------------------------------------------
\vspace{\skips}
\subsubsection{Memory of Previous Trials}
\vspace{\skips}
%----------------------------------------------------------------------------------------------------------------------------------------------------------------------------------

Next we consider possible memory effects. We define the quantity $\tilde{W} \equiv W - (3 + \bar{\epsilon} ) N$. Then $\langle \tilde{W} \rangle = 0$, regardless of what plan the hidden-variable theory uses. On the other hand, the hidden-variable theory can affect the standard deviation of $\tilde{W}$. If we denote by $\tilde{W}_0$ the value of $\tilde{W}$ obtained in the experiment, then the $p$-value we seek is the probability that the hidden-variable theory could have achieved $\tilde{W} \geq \tilde{W}_0$ by chance. To discuss an experiment in progress, we define
\begin{equation}
\tilde{W}_n \equiv \sum_{\alpha = 1}^n (W_\alpha - 3 - \bar{\epsilon} ) ,
\label{tildeWn}
\end{equation}
which is the contribution to $\tilde{W}$ after $n$ trials.

For sufficiently large $N$, we may assume that the probabilities are well approximated by a Gaussian probability distribution. Then we expect that as long as $\tilde{W}_n \leq \tilde{W}_0$, the best strategy for the hidden-variable theory is to maximize $\sigma_{\tilde{W}}$, so that the number of standard deviations to its goal is as small as possible. When and if $\tilde{W}_n$ passes $\tilde{W}_0$, on the other hand, then its best strategy is to minimize $\sigma_{\tilde{W}}$, so as to minimize the probability that $\tilde{W}$ might backslide to $\tilde{W} \leq \tilde{W}_0$. 

We define $N_{ {\rm lose}, ij}$ as the number of trials for which the hidden-variable theory selects settings $(ij)$ as the loser. Then we seek to estimate $p_{\rm left} (n \vert N_{ {\rm lose}, ij}) \equiv p (\tilde{W}_n < 0 )$, under the assumption that the hidden-variable loser selection is given by $N_{ {\rm lose} , ij}$. That is, $p_{\rm left} (n \vert N_{ {\rm lose} , ij})$ is the probability that after $n$ trials, the net change in $\tilde{W}$ has been to the left (i.e., negative). For large $n$, we expect the probability distribution for $\tilde{W}$ to become a Gaussian with zero mean, so that $p_{\rm left} (n \vert N_{ {\rm lose} , ij} )$ should approach $1/2$, for any hidden-variable theory loser selection. For smaller $n$, however, $p_{\rm left} (n \vert N_{ {\rm lose} , ij})$ can reach some maximum value $B > 1/2$.

%%%%%%%%%%%%%%%%%%%%%%%%%%%%%%%%%%%%%%%%
\begin{figure}[t]
\begin{center}
\includegraphics[width=0.45\textwidth]{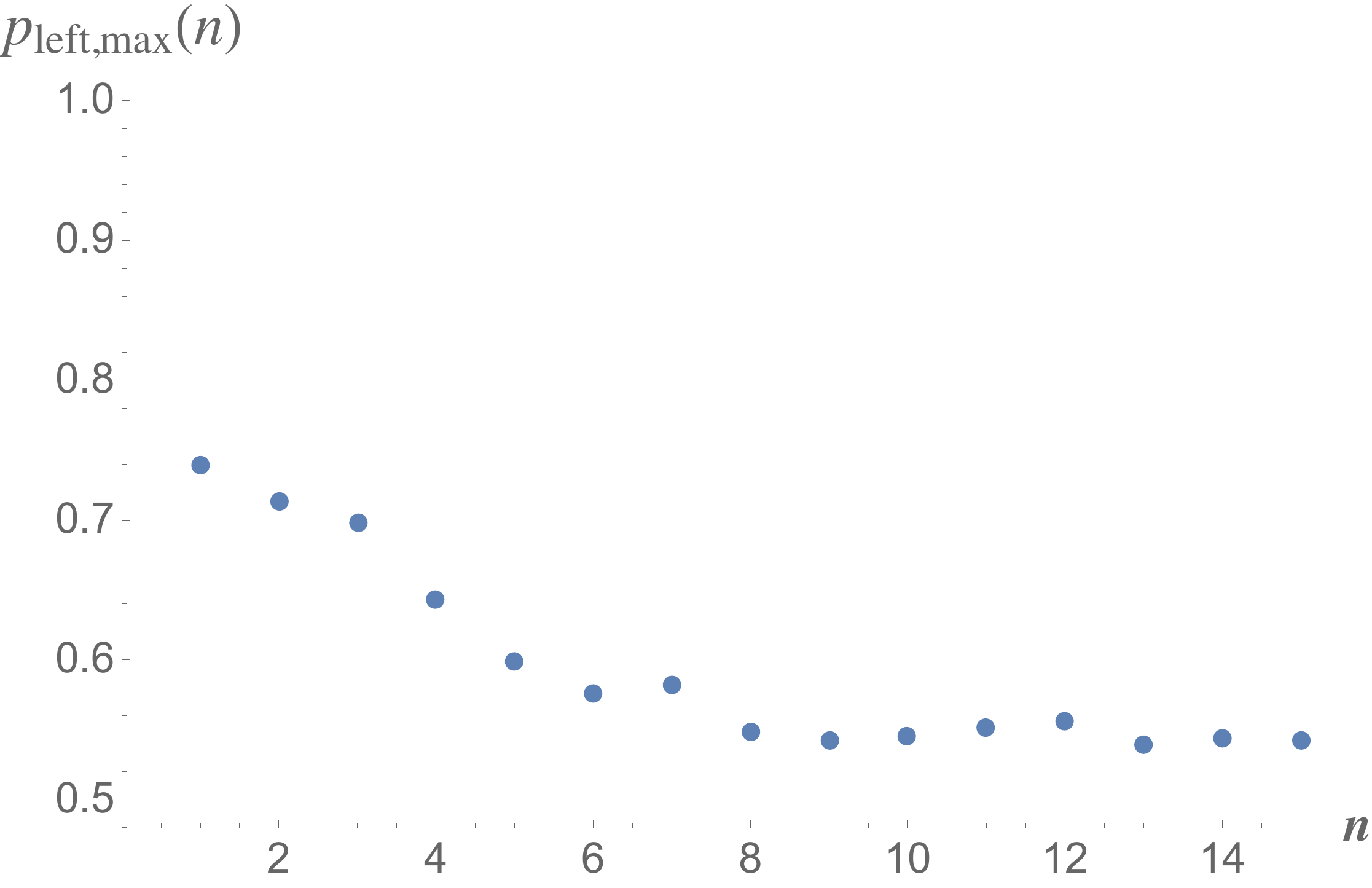}
\includegraphics[width=0.45\textwidth]{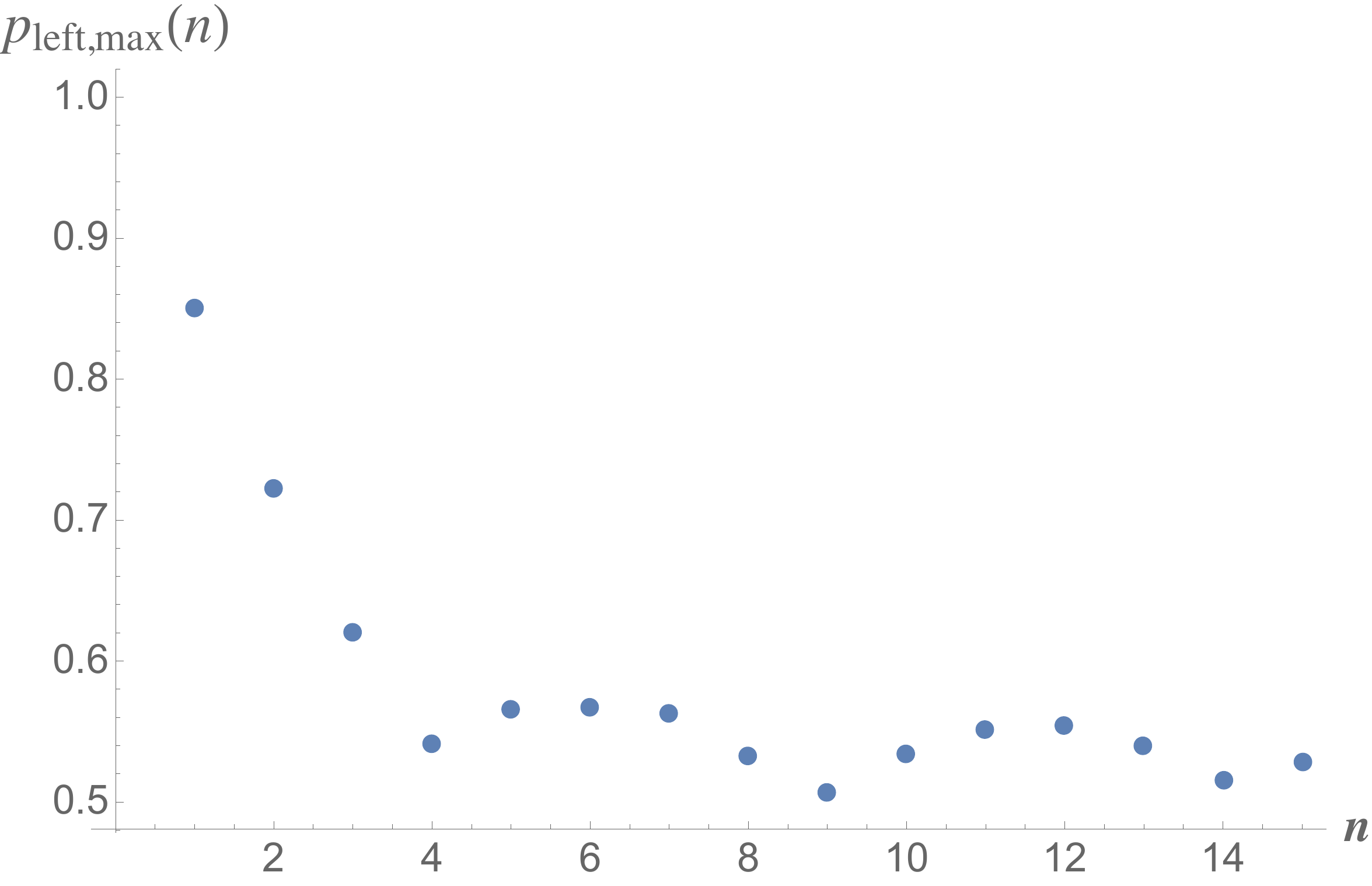}
\end{center}
\vspace{-0.6cm}
\caption{\small The quantity $p_{\rm left, max} (n)$ is the maximum probability that $\tilde{W}$ moves to the left in $n$ trials. Shown here is $p_{\rm left, max} (n)$ for experimental run 1 (top) and for run 2 (bottom).
}
\label{randomwalkfig}
\end{figure}
%%%%%%%%%%%%%%%%%%%%%%%%%%%%%%%%%%%%%%%%

Finally, we define $p_1$ to be the probability that $\tilde{W}_n \geq \tilde{W}_0$ for some $n$ in the range $1 \leq n \leq N$, under the assumption that the hidden-variable theory consistently makes choices that maximize $\sigma_{\tilde{W}}$. Consider some particular sequence of trials that contributes to $p_1$, that is, a sequence for which $\tilde{W}_n \geq \tilde{W}_0$ for some $n$. The continuation of this sequence for the rest of the experiment (assuming that the hidden-variable theory continues to make choices that maximize $\sigma_{\tilde{W}}$) can do one of two things: it can finish the experiment with $\tilde{W} \geq \tilde{W}_0$, or it can finish the experiment with $\tilde{W} < \tilde{W}_0$. In the first case, this sequence contributes to the $p$ value we calculated in the previous subsection, whereas in the second case it does not. The second case is an instance of backsliding, for which we know that the probability is at most $B$. Hence the probability of the first case is at least $1 - B$, so the $p$ value we seek, $p_{\rm mem}$, satisfies
\begin{equation}
p_{\rm mem} \leq \frac{ p}{1 - B } ,
\label{pmem}
\end{equation}
where $p$ is the value calculated in the previous subsection, which did not account for memory effects. Therefore it remains to estimate $B$.

The quantity $B = {\rm max}_n \{ p_{\rm left, max} (n) \}$, where $p_{\rm left, max} (n) \equiv {\rm max} \{ p_{\rm left} (n \vert N_{ {\rm lose}, ij} ) \}$, and the latter quantity is maximized over all possible assignments of (non-negative integers) $N_{ {\rm lose}, ij}$ consistent with $\sum_{ij} N_{ {\rm lose}, ij} = n$. Using the $\epsilon_{ij}$ and $q_{ij}$ for experimental runs 1 and 2, we find the results shown in Fig.~\ref{randomwalkfig}. For both experimental runs, the largest probability for a leftward excursion of $\tilde{W}$ occurred for $n = 1$. 
In particular, we find $B = 0.7393$ for run 1, and $B = 0.8500$ for run 2. 
Separate Monte Carlo simulations (involving 100 million samples) illustrate that $p_{\rm left} (n) \rightarrow 1/2$ for $n \sim 10^3$, considerably greater than the $n =15$ shown in Fig.~\ref{randomwalkfig}, but still much smaller than $N \sim 10^5$ in each experimental run.

Incorporating these memory effects, we find $p_{\rm mem} \leq 1.78 \times 10^{-13}$ for run 1, and $p_{\rm mem} \leq 3.96 \times 10^{-33}$ for run 2. Again assuming a Gaussian distribution, these correspond to $\nu \geq 7.31$ and $11.93$ standard deviations, respectively. 
We consider these numbers to be reasonable estimates of the statistical significance of our experimental results, deriving as they do from conservative assumptions applied to a calculation tailored specifically to our experimental setup.

%----------------------------------------------------------------------------------------------------------------------------------------------------------------------------------
\vspace{\skips}
\subsubsection{No-signaling}
\vspace{\skips}
%----------------------------------------------------------------------------------------------------------------------------------------------------------------------------------

Lastly, we check whether our data are consistent with the no-signaling principle. This principle demands that, under space-like separation, local outcome
probabilities must not depend on the setting of the distant party:%
\begin{equation}
\begin{split}
p(A\!  &  =\!+|a_{i}b_{j})=p(A\!=\!+|a_{i}b_{j^{\prime}}),\\
p(B\!  &  =\!+|a_{i}b_{j})=p(B\!=\!+|a_{i^{\prime}}b_{j}).
\label{pABnosignal}
\end{split}
\end{equation}
The analogous equations for the `$-$' outcomes follow trivially from $p(A\!=\!-|a_{i}b_{j})=1-p(A\!=\!+|a_{i}b_{j})$. Let us denote by $N_{a_{i}}^{\pm}$ ($N_{b_{j}}^{\pm}$)\ the number Alice's (Bob's) outcomes `$\pm$'
where she (he) had setting $a_{i}$ ($b_{j}$). The recorded data for experimental run 1, post-selecting only onto a valid setting choice (i.e., the click in the setting reader occurred within the time-interval $\tau_{\rm used}$) and not onto a coincident outcome at the distant location, were
\begin{equation}%
\begin{array}
[c]{lrrlrr}%
\vspace{0.1cm} & b_{1} & b_{2} &  & a_{1} & a_{2}\\
N_{a_{1}}^{+} & 163\,292 & 550\,046 & \;\;\;N_{b_{1}}^{+} & 562\,351 &
352\,896\\
N_{a_{2}}^{+} & 101\,289 & 340\,045 & \;\;\;N_{b_{2}}^{+} & 2\,033\,046 &
1\,279\,635\\
N_{a_{1}}^{-} & 165\,593 & 555\,034 & \;\;\;N_{b_{1}}^{-} & 480\,738 &
302\,277\\
N_{a_{2}}^{-} & 100\,848 & 340\,890 & \;\;\;N_{b_{2}}^{-} & 1\,553\,010 &
976\,740
\end{array}
\label{eq:nosignal}
\end{equation}
The data in Eq.~(\ref{eq:nosignal}) were obtained after applying the $\tau_{\rm cut}$ filter. We denote by $N_{a_{i},b_{j}}^{\pm}$ ($N_{b_{j},a_{i}}^{\pm}$) the value of $N_{a_{i}}^{\pm}$ ($N_{b_{j}}^{\pm}$) in the above table for distant setting $b_{j}$ ($a_{i}$). A point estimate for $p(A\!=\!+|a_{i}b_{j})$ is then given by $N_{a_{i},b_{j}}^{+}/(N_{a_{i},b_{j}}^{+}\!+\!N_{a_{i},b_{j}}^{-})$, and a point estimate for $p(B\!=\!+|a_{i}b_{j})$ is given by $N_{b_{j},a_{i}}^{+}/(N_{b_{j},a_{i}}^{+}\!+\!N_{b_{j},a_{i}}^{-})$.

Under space-like separation of all relevant events, no-signaling must be obeyed in both local realism and quantum mechanics, since its violation would contradict special relativity. (An experimental violation of no-signaling would require the settings of the distant laboratory to be available at the local measurement station via faster-than-light communication or due to a common cause in the past.) For run 1, point estimates yield the following probabilities: 
\begin{equation}
\begin{split}
p(A\!=\!+|a_{1}b_{1})  &  =0.4965,\;\;p(A\!=\!+|a_{1}b_{2})=0.4977,\\
p(A\!=\!+|a_{2}b_{1})  &  =0.5011,\;\;p(A\!=\!+|a_{2}b_{2})=0.4994,\\
p(B\!=\!+|a_{1}b_{1})  &  =0.5391,\;\;p(B\!=\!+|a_{2}b_{1})=0.5386,\\
p(B\!=\!+|a_{1}b_{2})  &  =0.5669,\;\;p(B\!=\!+|a_{2}b_{2})=0.5671.
\end{split}
\end{equation} 
The null hypothesis of no-signaling demands that the two conditional probabilities in each line should be equal. In order to test for signaling, we perform a pooled two-proportion $z$-test. The probabilities that the observed data or worse are obtained under the null hypothesis are 0.211, 0.177, 0.532, 0.654, respectively. (For the stars used in run 2, we obtain the probabilities 0.434, 0.342, 0.737, 0.582, respectively.) As all probabilities are large, our data are in agreement with the no-signaling assumption. (We performed the same test on our data for runs 1 and 2 prior to applying the $\tau_{\rm cut}$ filter, and likewise found no statistical evidence to suggest signaling.) 

We remark that when post-selecting on coincident outcome events, i.e.\ using the counts in Eq.~(\ref{eq coinc}), the condition $p(A\!=\!+|a_{2}b_{1},B\!=\!\ast)=p(A\!=\!+|a_{2}b_{2},B\!=\!\ast)$, where `$B\!=\!\ast$'
denotes that Bob had a definite outcome (whose value is ignored), is violated significantly in both experiments. This can be attributed to the fact that the two total detection efficiencies for outcomes `$+$' and `$-$,' especially on Bob's side, were not the same. Let us denote the total detection efficiencies of Alice (Bob) for outcome $\pm$ by $\eta_{\pm}^{(A)}$ ($\eta_{\pm}^{(B)}$), including all losses in the source, the link, and the
detectors themselves. A detailed quantum-mechanical model for the data of run 1 suggests that the ratios of these efficiencies were $R^{(A)}\equiv\eta_{-}^{(A)}/\eta_{+}^{(A)}= 1.00$ for Alice and $R^{(B)}\equiv\eta_{-}^{(B)}/\eta_{+}^{(B)}= 0.81$ for Bob. The difference in $R^{(A)}$ and $R^{(B)}$ can fully be understood on the basis of the known efficiency differences of the detectors used. One can correct the counts in Eq.~(\ref{eq coinc}) for these efficiencies by multiplying all `$+$' counts of Alice (Bob) by $\sqrt{R^{(A)}}$ ($\sqrt{R^{(B)}}$), and dividing all her (his) `$-$' counts by $\sqrt{R^{(A)}}$ ($\sqrt{R^{(B)}}$). The corrected counts show no sign of a violation of no-signaling. This is also true for the data from run 2. 

We finally note that, due to the low total detection efficiencies, our experiment had to make the assumptions of fair sampling and fair coincidences. This implies that low or imbalanced detection efficiencies are not exploited by hidden-variable models.


\begin{thebibliography}{74}
\bibitem [{Ret()}]{Retrofn}%
  \BibitemOpen
  \href@noop {} {}\bibinfo {note} {{One may even consider retrocausal models in
  which the relevant hidden variable affects the measurement settings from the
  future \cite{costabeauregard,pricewharton}. The point is that the relevant
  hidden variable may be associated with spacetime regions far removed from the
  source of entangled particles}}\BibitemShut {NoStop}%
\bibitem [{Sup()}]{Superdeterminism}%
  \BibitemOpen
  \href@noop {} {}\bibinfo {note} {{We note that it is only possible to fully
  close the locality and freedom-of-choice loopholes by making certain
  assumptions. If settings were fully determined at the beginning of the
  universe, they could always be known at the distant measurement locations or
  be correlated with the hidden variables.}}\BibitemShut {Stop}%
\end{thebibliography}
\end{document}